\documentclass[pra,twocolumn,showpacs,preprintnumbers,superscriptaddress,amsmath,amssymb]{revtex4}
\usepackage{graphicx}
\usepackage{subfigure}
\usepackage{mathrsfs}
\usepackage{amsfonts}
\usepackage{amsmath}
\usepackage{color}
\usepackage{multirow}
\usepackage[colorlinks,linkcolor=blue,citecolor=blue]{hyperref}

\begin{document}
\title{Quantum Trajectory Thermodynamics with Discrete Feedback Control}
\author{Zongping Gong}
\affiliation{Department of Physics, University of Tokyo, 7-3-1 Hongo, Bunkyo-ku, Tokyo 113-0033, Japan}
\author{Yuto Ashida}
\affiliation{Department of Physics, University of Tokyo, 7-3-1 Hongo, Bunkyo-ku, Tokyo 113-0033, Japan}
\author{Masahito Ueda}
\affiliation{Department of Physics, University of Tokyo, 7-3-1 Hongo, Bunkyo-ku, Tokyo 113-0033, Japan}
\affiliation{RIKEN Center for Emergent Matter Science (CEMS), Wako, Saitama 351-0198, Japan}
\date{\today}

\begin{abstract}
We employ the quantum jump trajectory approach to construct a systematic framework to study the thermodynamics at the trajectory level in a nonequilibrium open quantum system under discrete feedback control. Within this framework, we derive quantum versions of the generalized Jarzynski equalities, which are demonstrated in an isolated pseudospin system and a coherently driven two-level open quantum system. Due to quantum coherence and measurement backaction, a fundamental distinction from the classical generalized Jarzynski equalities emerges in the quantum versions, which is characterized by a large negative information gain reflecting genuinely quantum rare events. A possible experimental scheme to test our findings in superconducting qubits is discussed.
\end{abstract}
\maketitle

\section{Introduction} 
Recent years have witnessed the rise of an interdisciplinary field of information thermodynamics \cite{Sagawa2015}. Information processing and feedback control in small classical thermodynamic systems are fairly well understood in terms of thermodynamic variables \cite{Jarzynski1997,Crooks1998,Sekimoto1998,Seifert2005} and information gain \cite{Sagawa2010,Sagawa2012} along individual trajectories. In particular the generalized Jarzynski equalities \cite{Sagawa2010}
\begin{equation}
\langle e^{-\beta(W-\Delta F)}\rangle=\eta,\;\;\;\;\;\;\;\;
\langle e^{-\beta(W-\Delta F)-I}\rangle=1
\label{GJE}
\end{equation}
connect the work $W$ with the efficacy $\eta$ of feedback control and the mutual information $I$. Here $\Delta F$ and $\beta$ are respectively the free-energy difference and the inverse temperature. These relations have been experimentally verified by using colloidal particles \cite{Toyabe2010} and a single-electron box \cite{Pekola2014}.

However, there has been little progress in the quantum aspect of information thermodynamics at the trajectory level. The main difficulty is to identify the thermodynamic variables and the information content compatible with genuine quantum effects such as superposition and measurement backaction. The thermodynamics of information processing has been discussed mainly on the basis of statistical ensembles \cite{Sagawa2008,Sagawa2009,Funo2013b,Seifert2015,Pekola2016}, whereas only special cases have been examined at the trajectory level including classical measurement errors \cite{Tasaki2011}, an isolated driving \cite{Funo2013} and a  separated thermalization process \cite{Funo2015}. 

Meanwhile, there have been remarkable advances in experimental techniques to measure and control small quantum systems such as trapped ions \cite{Zoller2011}, quantum dots \cite{Taylor2012} and superconducting qubits \cite{Devoret2013}, which can be used to implement quantum information processing and operate in the presence of dissipation and dephasing. In particular, continuous monitoring 
\cite{Siddiqi2011,Siddiqi2013,Huard2016} and measurement-based feedback control \cite{DiCarlo2012,Devoret2013b} have been achieved in superconducting qubits. It thus seems timely to develop a theoretical framework to study quantum trajectory thermodynamics under feedback control.

Among various proposals for the definitions of work and heat in open \cite{Maes2004,Esposito2006,Esposito2009,Talkner2011} and isolated quantum systems \cite{Solinas2015,Talkner2015,Talkner2015b}, the quantum jump trajecotry (QJT) approach, which was originally developed in quantum optics \cite{Zoller1987,Dalibard1992,Plenio1998} and applied to quantum thermodynamics quite recently \cite{Crooks2008,Maes2008,Horowitz2012,Pekola2013,Horowitz2013,Breuer2013,Liu2014a,Liu2014b,Horowitz2015,Elouard2015,Nissila2015,Nissila2016}, provides a natural framework to define thermodynamic quantities. The QJT-based definition naturally incorporates quantum coherence and gives the definitions of work and heat that reduce to the widely accepted ones (see Appendix \ref{Consistency} for details) upon ensemble averaging \cite{Alicki1979,Kosloff1984} or in the classical \cite{Crooks1998} and adiabatic limits \cite{Kurchan2000,Tasaki2000,Talkner2007}. 

In this paper, we extend the QJT approach to a widely applicable quantum thermodynamic process with discrete feedback control to establish a framework for systematically studying information thermodynamics in small open quantum systems at the level of individual trajectories. Yet another genuinely quantum-mechanical effect -- measurement bakcaction -- is also included.  In particular, we find the quantum generalizations of Eq.~(\ref{GJE}) and highlight the fundamental distinction from their classical counterparts \cite{Sagawa2010}, which is characterized by a new information content (\ref{IG}) that signals \emph{quantum rare events} by large negative values. 
The present work thus significantly broadens the scope of information thermodynamics to open quantum systems, where quantum coherent thermodynamics, measurement backaction, and feedback control may conspire to yield as yet unexplored emergent quantum phenomena.

This paper is structured as follows. In Sec.~\ref{QTD}, we review the quantum master equation formalism of quantum thermodynamics at the ensemble level. In Sec.~\ref{QTTD}, we review the quantum trajectory thermodynamics in the absence of feedback control. In Sec.~\ref{TQFCP}, we combine quantum trajectory thermodynamics with feedback control to establish the general framework for information thermodynamics in the quantum regime. We derive the quantum versions of the generalized Jarzynski equalities in Sec.~\ref{GQJE}. Two examples are given in Sec.~\ref{ex}. Finally we conclude the paper in Sec.~\ref{conclu}. Several complicated algebraic manipulations and detailed discussions are relegated to appendices to avoid digressing from the main subject. Appendix \ref{apdA} provides a detailed derivation of the master equation (\ref{EOM}). Appendix \ref{RQTT} shows how heat and work can be defined without ambiguity along a single quantum trajectory. Appendix \ref{Consistency} demonstrates how the QJT-based definitions of work and heat reduce to their widely accepted definitions at the ensemble level and in the classical or adiabatic limit. Appendix \ref{DDGJE} gives derivations of the generalized quantum Jarzynski equations. Appendix \ref{Example} describes some details of the example discussed in Sec. \ref{ex}.

\section{Quantum thermodynamics}
\label{QTD} 
\subsection{Markovian quantum master equation}
\begin{figure}
\begin{center}
        \includegraphics[width=8.5cm, clip]{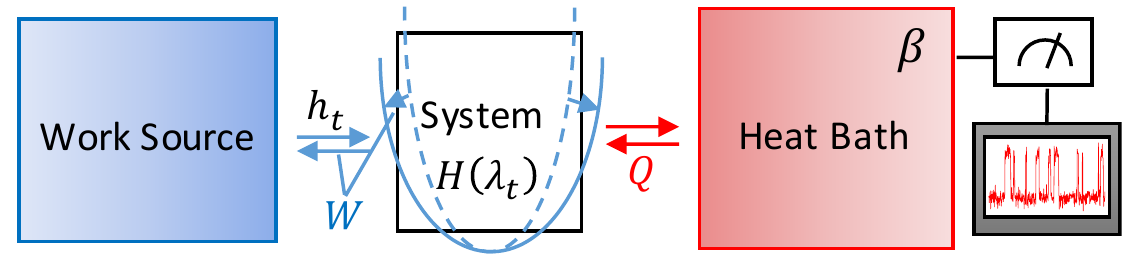}\\
      \end{center}
   \caption{(color online). 
   A system is weakly coupled to an ideal heat bath with inverse temperature $\beta$ and simultaneously driven out of equilibrium by a time-dependent inclusive Hamiltonian $H(\lambda_t)$ and an exclusive one $h_t$. At the trajectory level, the system is projectively measured twice under the eigen basis of the intantaneous Hamiltonian at the initial and final times, while the heat bath is under continuous projective monitoring during the whole process.}
   \label{fig1}
\end{figure}

We consider a $d$-level system with nondegenerate energy gaps, whose state at time $t$ is described by the density operator $\rho_t$. As schematically illustrated by Fig.~\ref{fig1}, the system is under nonequilibrium driving and weakly coupled to a large heat bath at inverse temperature $\beta$. The time-dependent driving can be classified into an \emph{inclusive} part $H(\lambda_t)$ with a tunable work parameter $\lambda_t$ and an \emph{exclusive} part $h_t$ \cite{Jarzynski2007}, where only the former is included in the system energy $E_t=\mathrm{Tr}[\rho_t H(\lambda_t)]$ while the latter arises from external driving. We assume a sufficiently slow inclusive driving speed $\dot\lambda_t$ and a short memory time $\tau_{\rm B}$ of the heat bath \cite{Lidar2012} (see Eq.~(\ref{judgement}) for details). Under the Born-Markov approximation and the rotating-wave approximation \cite{Breuer2002,Wiseman2010}, the Lindblad master equation \cite{Lindblad1976} can be obtained as (see Appendix \ref{apdA} for the derivation) 
\begin{equation}
\dot\rho_t=\mathcal{L}_t\rho_t=-\frac{i}{\hbar}[H(\lambda_t)+h_t,\rho_t]+\sum_j\mathcal{D}[L_j(\lambda_t)]\rho_t,
\label{EOM}
\end{equation}
where $\mathcal{D}[c]\rho\equiv c\rho c^\dag-\{c^\dag c,\rho\}/2$ is a traceless superoperator, and $L_j(\lambda)$ is the $j$-th jump operator satisfying $[L_j(\lambda),H(\lambda)]=\Delta_j(\lambda)L_j(\lambda)$ with $\Delta_j(\lambda)\in\{E^\lambda_k-E^\lambda_l:H(\lambda)|k^\lambda\rangle=E^\lambda_k|k^\lambda\rangle,\;k,l=1,2,...,d\}$ and the detailed balance condition $L^\dag_{j'}(\lambda)=L_j(\lambda)e^{-\beta\Delta_j(\lambda)/2}$ with $j'$ uniquely determined from $\Delta_{j'}(\lambda)=-\Delta_j(\lambda)$ if $\Delta_j(\lambda)\neq0$ and $j'=j$ otherwise. The Lamb shift is ignored. Since $h_t$ is exlusive, the detailed balance condition ensures the system to relax to an instantaneous equilibrium state only when $\lambda_t$ is constant and 
$h_t$ is turned off \cite{Liu2014a}. 

While we introduce Eq.~(\ref{EOM}) based on the standard ``small system $+$ large environment" approach \cite{Pekola2013}, the same equation of motion can be obtained for an effective heat bath constituted from a set of independent and identically distributed systems (e.g., two-level atoms \cite{Horowitz2012}), each of which that sequentially interacts with the system during an appropriate short time \cite{Horowitz2013}. This can be understood from the fact that any Markovian, completely positive and trace-preserving (CPTP) open quantum dynamics possesses the Lindblad form \cite{Breuer2004}. Equation (\ref{EOM}) is valid if $h_t$ is perturbative (i.e., $h_t\ll H(\lambda_t)$) or represents a sequence of sudden pulses (see Appendix \ref{apdA} for a heuristic argument). 
Thus, our formalism applies to a broad class of driving protocols such as $\pi$-pulses used in Ref.~\cite{Pekola2016} and potentially to quantum computation \cite{Nielsen2010}, where a gate operation $U_g=e^{-ih_g}$ at time $t_g$ can be generated by $h_t=\hbar h_g\delta(t-t_g)$. 

\subsection{Work and heat at the ensemble level}

In the absence of an exclusive driving ($h_t=0$), we have the following well-known expressions for work and heat \cite{Alicki1979,Kosloff1984}: 
\begin{equation}
\begin{split}
\langle W\rangle&=\int^\tau_0 dt\dot\lambda_t\mathrm{Tr}[\partial_\lambda H(\lambda_t)\rho_t],\\
\langle Q\rangle&=-\int^\tau_0 dt\mathrm{Tr}[H(\lambda_t)\dot\rho_t],
\label{inQW}
\end{split}
\end{equation}
where $\rho_t$ is the solution to Eq.~(\ref{EOM}). 
Such definitions allow intuitive interpretations that the energy change due to a change of the work parameter (the state) is identified as work (heat), and satisfy the first law of thermodynamics $\Delta E\equiv E_\tau-E_0=\langle W\rangle-\langle Q\rangle$. Here $Q>0$ corresponds to the heat transferred \emph{from the system to the heat bath}. 

In the presence of an exclusive driving ($h_t\neq0$), the expression of work and heat should be modified by
\begin{equation}
\begin{split}
\langle W\rangle=&\int^\tau_0 dt\dot\lambda_t\mathrm{Tr}[\partial_\lambda H(\lambda_t)\rho_t]\\
&-\frac{1}{i\hbar}\int^\tau_0 dt\mathrm{Tr}[[h_t,H(\lambda_t)]\rho_t],\\
\langle Q\rangle=&-\int^\tau_0 dt\mathrm{Tr}[H(\lambda_t)\dot\rho_t]\\
&-\frac{1}{i\hbar}\int^\tau_0 dt\mathrm{Tr}[[h_t,H(\lambda_t)]\rho_t].
\end{split}
\label{QWQ}
\end{equation}
Here, additional terms appears due to the fact that $h_t$ affects the unitary part of the dynamics just like $H(\lambda_t)$, but the effect is excluded when we evaluate the energy expectation. If we used Eq.~(\ref{inQW}), for a short time interval $[t,t+dt]$, an additional energy change $-\frac{i}{\hbar}{\rm Tr}[[h_t,\rho_t]H(\lambda_t)]dt=-\frac{1}{i\hbar}{\rm Tr}[[h_t,H(\lambda_t)]\rho_t]dt$ due to the unitary state evolution contributed by $h_t$ would be misidentified as heat. 

A simple illustrative example is a situation relevant to the quantum Bochkov-Kuzovlev equalities for isolated systems \cite{Campisi2011}, where $\lambda_t=\lambda, \forall t\in[0,\tau]$ is fixed so that $\langle W\rangle=\Delta E={\rm Tr}[H(\lambda)(\rho_\tau-\rho_0)]$ and $\langle Q\rangle=0$. One can check that Eq.~(\ref{QWQ}) indeed gives this result, while Eq.~(\ref{inQW}) leads to the wrong results: $\langle Q\rangle={\rm Tr}[H(\lambda)(\rho_\tau-\rho_0)]$ and $\langle W\rangle=0$. An interesting special limit  is the quantum logic gate operation with $h_t=\hbar h_g\delta(t-t_g)$. Suppose that the input state is $\rho_{t^-_g}$, the energy cost, which is attributed to work, of the quantum logic gate operation $U_g=e^{-ih_g}$ generated by $h_t$ should be $\langle W_g\rangle =\mathrm{Tr}[H(\lambda)(\rho_{t^+_g}-\rho_{t^-_g})]$, where $\rho_{t^+_g}=U_g\rho_{t^-_g}U^\dag_g$ is the quantum state after the operation. It is clear that Eq.~(\ref{QWQ}) gives such an result. However, if we used Eq.~(\ref{inQW}), we would again arrive at a wrong conclusion that such an energy cost is identified as heat.


To further convince ourselves the necessity of the additional terms in Eq.~(\ref{QWQ}), we may recall the classical counterpart. 
As is well known in classical stochastic thermodynamics, the work functional with respect to a trajectory $\Gamma_t$ in the phase space of a Brownian particle with mass $M$, subjected to a nonconservative force $f_t$ and confined in a time-dependent potential $V(x,\lambda_t)$, is \cite{Seifert2008}
\begin{equation}
W_{\rm C}[\Gamma_t]=\int^\tau_0 dt\dot\lambda_t\partial_\lambda V(x_t,\lambda_t)+\frac{1}{M}\int^\tau_0 dt p_tf_t.
\end{equation}
Suppose that $f_t$ arises from a fictitious potential $h_t(x)\equiv-f_tx$. Then $W_{\rm C}[\Gamma_t]$ can be rewritten as
\begin{equation}
\begin{split}
W_{\rm C}[\Gamma_t]&=\int^\tau_0 dt\dot\lambda_t\partial_\lambda H(x_t,\lambda_t)\\
&-\int^\tau_0 dt \{h_t(x_t),H(x_t,\lambda_t)\}_{\rm PB},
\end{split}
\end{equation}
where $H(x,\lambda)=p^2/2M+V(x,\lambda)$ is the classical Hamiltonian, and $\{\cdot,\cdot\}_{\rm PB}$ is the Poisson bracket. By replacing the Poisson bracket with $\frac{1}{i\hbar}[\cdot,\cdot]$ and taking the ensemble average $\mathrm{Tr}[\rho_t...]$, we obtain Eq.~(\ref{QWQ}).

\section{Quantum trajectory thermodynamics} 
\label{QTTD}
\subsection{Quantum jump trajectory}
While classical trajectory thermodynamics or stochastic thermodynamics is a relatively mature field \cite{Broeck2015}, few progresses have been made on its quantum generalization until very recent years (see Appendix \ref{RQTT} for some useful remarks). Interestingly, this cutting-edge problem is found to be closely related to the well-established QJT approach, which we briefly review here.

According to the equation of motion (\ref{EOM}), up to accuracy $O(\delta t^2)$, $\rho_{t+\delta t}$ can be expressed as the nonselective postmeasurement state of $\rho_t$ with respect to a certain measurement \cite{Horowitz2015}:
\begin{equation}
\begin{split}
\rho_{t+\delta t}&=\left[I-\frac{i}{\hbar}H_{\rm eff}(t)\delta t\right]\rho_t\left[I+\frac{i}{\hbar}H^\dag_{\rm eff}(t)\delta t\right]\\
&+\sum_j L_j(\lambda_t)\sqrt{\delta t}\rho_tL^\dag_j(\lambda_t)\sqrt{\delta t},
\end{split}
\end{equation}
where $H_{\rm eff}(t)=H(\lambda_t)+h_t-\sum_ji\hbar L^\dag_j(\lambda_t)L_j(\lambda_t)/2$ is the \emph{non-Hermitian} effective Hamiltonian. In a selective manner, we can interpret the open quantum dynamics during a short time interval as follows: there is a probability $\delta p_j=\mathrm{Tr}[L^\dag_j(\lambda_t)L_j(\lambda_t)\rho_t]\delta t$ ($p_0=1-\sum_j\mathrm{Tr}[L^\dag_j(\lambda_t)L_j(\lambda_t)\rho_t]\delta t$) of the outcome $j\neq0$ (0) being observed, which is accompanied by the backaction that changes $\rho_t$ into $L_j(\lambda_t)\rho_tL^\dag_j(\lambda_t)\delta t/\delta p_j$ ($\left[I-\frac{i}{\hbar}H_{\rm eff}(t)\delta t\right]\rho_t\left[I+\frac{i}{\hbar}H^\dag_{\rm eff}(t)\delta t\right]/p_0$). If $\rho_t$ is a pure state $|\psi_t\rangle\langle\psi_t|$, it will stay pure but differs for different outcomes. In particular, if the outcome $j=0$ is observed, we have
\begin{equation}
\begin{split}
&|\psi_{t+\delta t}\rangle=\frac{1}{\sqrt{p_0}}\left[I-\frac{i}{\hbar}H_{\rm eff}(t)\delta t\right]|\psi_t\rangle\\
&=\left[I-\frac{i}{\hbar}H_{\rm eff}(t)\delta t+\frac{1}{2}\sum_j\|L_j(\lambda_t)|\psi_t\rangle\|^2\delta t\right]|\psi_t\rangle,
\end{split}
\end{equation}
which describes a state change of the order of $O(\delta t)$ called \emph{nonunitary evolution}. If $j\neq0$ is observed, we have
\begin{equation}
|\psi_{t+\delta t}\rangle=\sqrt{\frac{\delta t}{\delta p_j}}L_j(\lambda_t)|\psi_t\rangle=\frac{L_j(\lambda_t)|\psi_t\rangle}{\|L_j(\lambda_t)|\psi_t\rangle\|},
\end{equation}
which describes a state change of the order of $O(1)$ due to a \emph{quantum jump} (QJ). Combining these two different types of evolutions, we obtain the QJ-type stochastic Schr\"odinger equation:
\begin{equation}
\begin{split}
d|\psi_t\rangle&=\left[-\frac{i}{\hbar}H_{\rm eff}(t)+\frac{1}{2}\sum_j\|L_j(\lambda_t)|\psi_t\rangle\|^2\right]|\psi_t\rangle dt\\
&+\sum_j\left[\frac{L_j(\lambda_t)}{\|L_j(\lambda_t)|\psi_t\rangle\|^2}-I\right]|\psi_t\rangle dN^j_t,
\end{split}
\label{SSE}
\end{equation}
where $dN^j_t$'s are independent random variables satisfying $(dN^j_t)^2=dN^j_t$ and $\mathrm{E}[dN^j_t]=\|L_j(\lambda_t)|\psi_t\rangle\|^2dt$. This stochastic Schr\"odinger equation is known as an \emph{unravelling} of the original LME (\ref{EOM}), in the sense that $\rho_t$ can be reproduced by taking the average over all the possible realizations of $|\psi_t\rangle$, i.e., $\rho_t=\mathrm{E}[|\psi_t\rangle\langle\psi_t|]$. It is worth mentioning that the unravelling is not unique. For the same LME, we also have the \emph{quantum-state diffusion} unravelling \cite{Gisin1992} in addition to the QJ-type one.

\subsection{Work and heat at the trajectory level}

While the QJ-type stochastic Schr\"odinger equation (\ref{SSE}) was originally  proposed for numerical computations \cite{Dalibard1992}, its physical interpretation was soon found in a specific direct photondetection process \cite{Wiseman1993}. Here the photon field serves as the heat bath (though it is the zero-temperature vacuum in Ref.~\cite{Wiseman1993}). Thus, the interpretation can be straightforwardly generalized to the continuous projective monitoring of the heat bath (see Fig.~\ref{fig1}). In the context of quantum thermodynamics, such an idea was first discussed in Ref.~\cite{Crooks2008}. 


The QJT approach presupposes a pure initial state. This condition is achieved by a projective measurement (PM) under the eigen basis of the initial Hamiltonian $H(\lambda_0)$; the PM also determines the initial energy $E^{\lambda_0}_a$ with $a$ being some quantum number. Furthermore, we perform another PM under the eigen basis of $H(\lambda_\tau)$ at the final time, which determines the final energy $E^{\lambda_\tau}_b$ (Fig.~\ref{fig1}). This two-time energy measurement (TTEM) scheme is inherited from the well-investigated special cases for isolated quantum systems \cite{Kurchan2000,Tasaki2000}. It is worth mentioning that the TTEM scheme is applicable only if $[\rho_0,H(\lambda_0)]=0$. Fortunately, this condition is satisfied if $\rho_0$ is the canonical distribution, which is the case in this paper. Generalization to a coherent initial state ($[\rho_0,H(\lambda_0)]\neq0$) remains an open problem \cite{Solinas2015}.

Each QJT $\psi_t$ represents a single individual realization of Eq.~(\ref{SSE}), with definite $dN^j_t$ and quantum number $a$ ($b$) obtained by continuously monitoring the heat bath and the initial (final) PM. Practically, a QJT corresponds to a sequence of outcomes observed in a \emph{single-shot} experiment. In terms of single-shot readouts, the heat $Q[\psi_t]$ and work $W[\psi_t]$ along such a QJT can be evaluated as \cite{Horowitz2012,Horowitz2013,Liu2014a,Liu2014b}
\begin{equation}
\begin{split}
Q[\psi_t]&=\sum_j\int^\tau_0 dN^j_t\Delta_j(\lambda_t),\\
W[\psi_t]&=E^{\lambda_\tau}_b-E^{\lambda_0}_a+\sum_j\int^\tau_0 dN^j_t\Delta_j(\lambda_t).
\end{split}
\label{heatwork}
\end{equation}
We can see that once the $j$-th QJ occurs at time $t$, the accumulated heat increases by $\Delta_j(\lambda_t)$ (which may be negative), so the heat is ``counted" discretely at the trajectory level. Combining the heat with the energy change determined by the initial and final PM outcomes, the work can be found from the first law of thermodynamics at the trajectory level.

\begin{figure}
\begin{center}
        \includegraphics[width=8.5cm, clip]{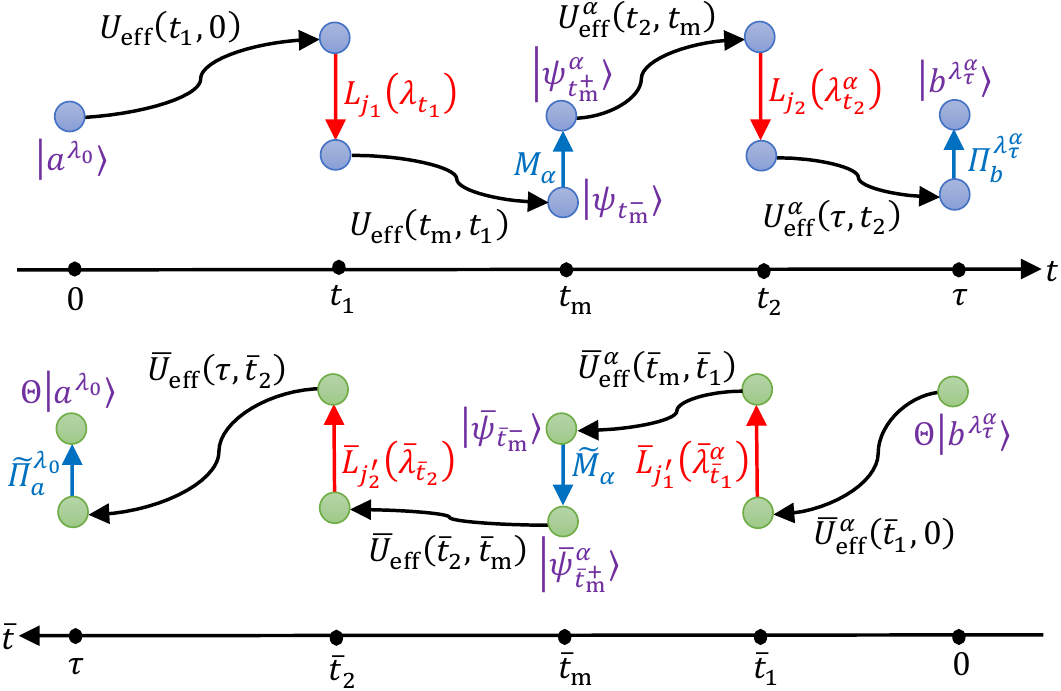}\\
      \end{center}
   \caption{(color online). 
   A forward QJT (upper panel) and the corresponding time-reversed QJT (lower panel) starting from the initial energy eigenstates $|a^{\lambda_0}\rangle$ and $\Theta|b^{\lambda^\alpha_\tau}\rangle$, respectively, where $\Theta$ is the time-reversal operator. Here the $j_1$-th ($j'_1$-th) quantum jump occurs at $t_1<t_{\mathrm{m}}$ ($\bar t_1<\bar t_{\mathrm{m}}$), with the outcome of $\mathrm{M_A}$ ($\mathrm{M_{B_\alpha}}$) being $\alpha$, the $j_2$-th ($j'_2$-th) quantum jump occurs at $t_2>t_{\mathrm{m}}$ ($\bar t_2>\bar t_{\mathrm{m}}$), and the forward (backward) trajectory ends at the final energy eigenstate $|b^{\lambda^\alpha_\tau}\rangle$ ($\Theta|a^{\lambda_0}\rangle$) due to the projective measurement $\Pi^{\lambda^\alpha_\tau}$ ($\tilde\Pi^{\lambda_0}\equiv\Theta\Pi^{\lambda_0}\Theta^\dag$). In general, $|\bar\psi_{\tau-t}\rangle\neq\Theta|\psi_t\rangle$.}
   \label{fig2}
\end{figure}


\section{Quantum feedback control}
\label{TQFCP}
\subsection{Discrete feedback control}
We are now in a position to apply quantum trajectory thermodynamics to feedback control, which is the main object of this paper. Complementary to continuous feedback controls \cite{Wiseman1994,Esposito2013,Lutz2016}, we consider the following measurement-based (discrete) feedback control \cite{Sagawa2008}. (i) Initially ($t=0$), the system is at thermal equilibrium, i.e., $\rho_0=e^{-\beta H(\lambda_0)}/Z^{\lambda_0}$ with $Z^\lambda\equiv\mathrm{Tr}[e^{-\beta H(\lambda)}]$. A PM $\Pi^{\lambda_0}$ is performed to determine the initial energy of the system, where $\Pi^\lambda\equiv\{|k^\lambda\rangle\langle k^\lambda|:H(\lambda)|k^\lambda\rangle=E^\lambda_k|k^\lambda\rangle,k=1,2,...,d\}$. (ii) During $0<t<t_{\mathrm{m}}$, the system evolves under a fixed protocol $\lambda_t$ and $h_t$. (iii) At $t=t_{\mathrm{m}}$, a general measurement described by a set of measurement operators $\mathrm{M_A}\equiv\{M_\alpha:\alpha\in\mathrm{A}\}$ with $\sum_\alpha M^\dag_\alpha M_\alpha=I$ ($I$ is the identity operator) is performed on the system. We assume that the measurement device is initialized to be in a pure state and that the measurement time is negligible. (iv) During $t_{\mathrm{m}}<t<\tau$, we choose driving protocols $\lambda^\alpha_t$ and $h^\alpha_t$ that depend on measurement outcomes $\alpha$. (v) Finally, at $t=\tau$, a PM $\Pi^{\lambda_\tau}$ is performed to determine the final energy of the system. 

\subsection{Work and heat in feedback control processes}
In the presence of feedback control, a QJT can be constructed as follows: (i) Starting from an energy eigenstate  $|a^{\lambda_0}\rangle$, the system's state $|\psi_t\rangle$ evolves stochastically according to Eq.~(\ref{SSE}) with fixed $\lambda_t$ and $h_t$. (ii) Conditioned on the system's state $|\psi_{t^-_\mathrm{m}}\rangle$ just before the measurement, there is a probability $\|M_\alpha|\psi_{t^-_\mathrm{m}}\rangle\|^2$ to observe an outcome $\alpha$, which entails a sudden state change into $|\psi^\alpha_{t^+_\mathrm{m}}\rangle=M_\alpha|\psi_{t^-_\mathrm{m}}\rangle/\|M_\alpha|\psi_{t^-_\mathrm{m}}\rangle\|$ due to the measurement backaction. (iii) The system evolves stochastically according to Eq.~(\ref{SSE}) with driving protocol $\lambda^\alpha_t$ and $h^\alpha_t$, and finally ends at $|b^{\lambda^\alpha_\tau}\rangle$ after the second PM. A typical QJT is schematically illustrated in Fig.~\ref{fig2} (upper half). 

By identifying the energy cost of the measurement as work \cite{Sagawa2008}, the heat and work along a QJT are evaluted by Eq.~(\ref{heatwork}) with $\lambda_t$ replaced by $\lambda^\alpha_t$ for $t>t_{\rm m}$. By defining $\lambda^\alpha_t\equiv\lambda_t$ ($\forall\alpha\in\mathrm{A}$) for $t<t_{\mathrm{m}}$ for convenience, we have
\begin{equation}
\begin{split}
Q[\psi_t,\alpha]&=\sum_j\int^\tau_0 dN^j_t\Delta_j(\lambda^\alpha_t),\\
W[\psi_t,\alpha]&=E^{\lambda^\alpha_\tau}_b-E^{\lambda_0}_a+\sum_j\int^\tau_0 dN^j_t\Delta_j(\lambda^\alpha_t).
\end{split}
\end{equation}

\section{Generalized quantum Jarzynski equalities} 
\label{GQJE}
While the fluctuation patterns of work and heat can be rather complex owing to the restriction on the dynamics imposed by the detailed balance condition, the fluctuations share some universal properties, which are captured by the fluctuation theorems \cite{Esposito2009,Talkner2011}. In the presence of feedback control, by adding certain correction terms due to measurement \cite{Tasaki2011,Funo2013,Funo2015}, we can derive some generalized fluctuation theorems. 

A simple derivation of the fluctuation theorems is to invoke the time-reversed (TR) process. Due to the measurement backaction, in the TR process for a given $\alpha$ we should not only reverse the driving protocol, but also perform a measurement $\mathrm{M_{B_\alpha}}$ at $\bar t_{\mathrm{m}}\equiv\tau-t_{\mathrm{m}}$, where $\tilde M_\alpha\equiv\Theta M^\dag_\alpha\Theta^\dag\in\mathrm{M_{B_\alpha}}$ ($\alpha\in\mathrm{B}_\alpha$) with $\Theta$ being the time-reversal operator. The other measurement operators in $\mathrm{M_{B_\alpha}}$ can be arbitrary since we only postselect the TR QJTs with outcome $\alpha$. Then for a given measurement outcome $\alpha$, the TR dynamics for $t\neq \bar t_{\rm m}$ is described by
\begin{equation}
\dot\rho_t=\mathcal{\bar L}^\alpha_t\rho_t=-\frac{i}{\hbar}[\bar H(\bar\lambda^\alpha_t)+\bar h^\alpha_t,\rho_t]+\sum_j\mathcal{D}[\bar L_j(\bar\lambda^\alpha_t)]\rho_t,
\label{TREOM}
\end{equation}
where $\bar\lambda^\alpha_t\equiv\lambda^\alpha_{\tau-t}$ and $\bar O_t=\Theta O_{\tau-t}\Theta^\dag$ if the operator is explicitly time-dependent and $\bar O=\Theta O\Theta^\dag$ otherwise. Consequently, we find the following trajectory version of the detailed balance condition (see Appendix \ref{DTDB}):
\begin{equation}
\mathcal{P}[\psi_t,\alpha]=e^{\beta(W[\psi_t,\alpha]-\Delta F_\alpha)}\mathcal{\bar P}[\bar\psi_t,\alpha],
\label{TrajDB}
\end{equation}
where $\mathcal{P}[\psi_t,\alpha]$ ($\mathcal{\bar P}[\bar\psi_t,\alpha]$) is the probability of a forward (TR) QJT with the total of $K$ QJs associated with $L_{j_k}(\lambda^\alpha_t)$ ($\bar L_{j'_k}(\bar\lambda^\alpha_t)$) at $t_k$ ($\bar t_k\equiv\tau-t_{K+1-k}$) and the measurement outcome of $\mathrm{M_A}$ ($\mathrm{M_{B_\alpha}}$) being $\alpha$, and $\Delta F_\alpha=\beta^{-1}\ln(Z^{\lambda_0}/Z^{\lambda^\alpha_\tau})$ is the free-energy difference. A typical TR QJT is presented in Fig.~\ref{fig2} (lower half).

\subsection{First main result}
Based on Eq.~(\ref{TrajDB}), we can derive the quantum versions of Eq.~(\ref{GJE}). The efficacy of feedback control reads (see Appendix \ref{DEFC} for the derivation)
\begin{equation}
\eta_{\rm{QJT}}=\sum_\alpha\mathrm{Tr}[\tilde M^\dag_\alpha\tilde M_\alpha\bar\rho^\alpha_{\bar t^-_\mathrm{m}}], 
\label{EFC}
\end{equation}
where $\bar t^-_{\mathrm{m}}\equiv\tau-t^+_{\mathrm{m}}$ and $\bar\rho^\alpha_t$ is the solution to the TR Lindblad quantum master equation (\ref{TREOM}) starting from the canonical ensemble $e^{-\beta\bar H(\lambda_\tau)}/Z^{\lambda_\tau}$. The classical result can be reproduced for a general classical measurement $M_\alpha=\sum_n \sqrt{p_{\alpha|n}}|n\rangle\langle n|$ (which is always Hermitian) with $\mathrm{A}=\{1,2,...,d\}$, where $n$ labels the classical states. In general, however, we should distinguish $M_\alpha$ from $\tilde M_\alpha$. The quantum Jarzynski equality can also be reproduced by setting $|A|=1$, i.e., $A=\{\alpha\}$ contains only a single outcome, and $M_\alpha=I$, which leads to $\eta_{\rm QJT}=1$.

A simple but important corollary of Eq.~(\ref{EFC}) is that $\langle e^{-\beta(W-\Delta F)}\rangle=\eta_{\rm QJT}\le\sum_\alpha{\rm Tr}[\tilde M^\dag_\alpha\tilde M_\alpha]={\rm Tr}[I]=d$. By using Jensen's inequality $\langle e^{x}\rangle\ge e^{\langle x\rangle}$, we obtain
\begin{equation}
-\langle W\rangle\le -\langle\Delta F\rangle+\beta^{-1}\ln d.
\end{equation}
This inequality implies that the ultimate limit of the extractable work in a quantum feedback control process \emph{cannot} exceed the classical one (notice that the Landauer bound \cite{Landauer1961} corresponds to the special case with $\Delta F=0$ and $d=2$). We note that a similar ``negative" conclusion has been drawn for the efficiency of quantum Carnot engines \cite{Deffner2015}. However, quantum enhancement of thermodynamic performance does exist for finite-time processes \cite{Seifert2015}.

Experimentally, $\eta_{\rm QJT}$ can be measured as follows: (i) for a fixed TR driving associated with $\alpha$ and from the equilibrium state, we statistically estimate the probability to observe outcome $\alpha$ for the measurement $\mathrm{M_{B_\alpha}}$ performed at $\bar t_{\mathrm{m}}$, and denote the obtained probability by $\tilde p_\alpha$ after repeating the same process many times; (ii) we change the TR driving, estimate $\tilde p_\alpha$ for all $\alpha\in\mathrm{A}$, and finally sum them up. One can see that such a scheme does not require any knowledge about the details of the microscopic dynamics. This observation is very similar to the classical counterpart \cite{Sagawa2010,Toyabe2010}.

\subsection{Second main result}

\begin{figure}
\begin{center}
        \includegraphics[height=2.62cm, clip]{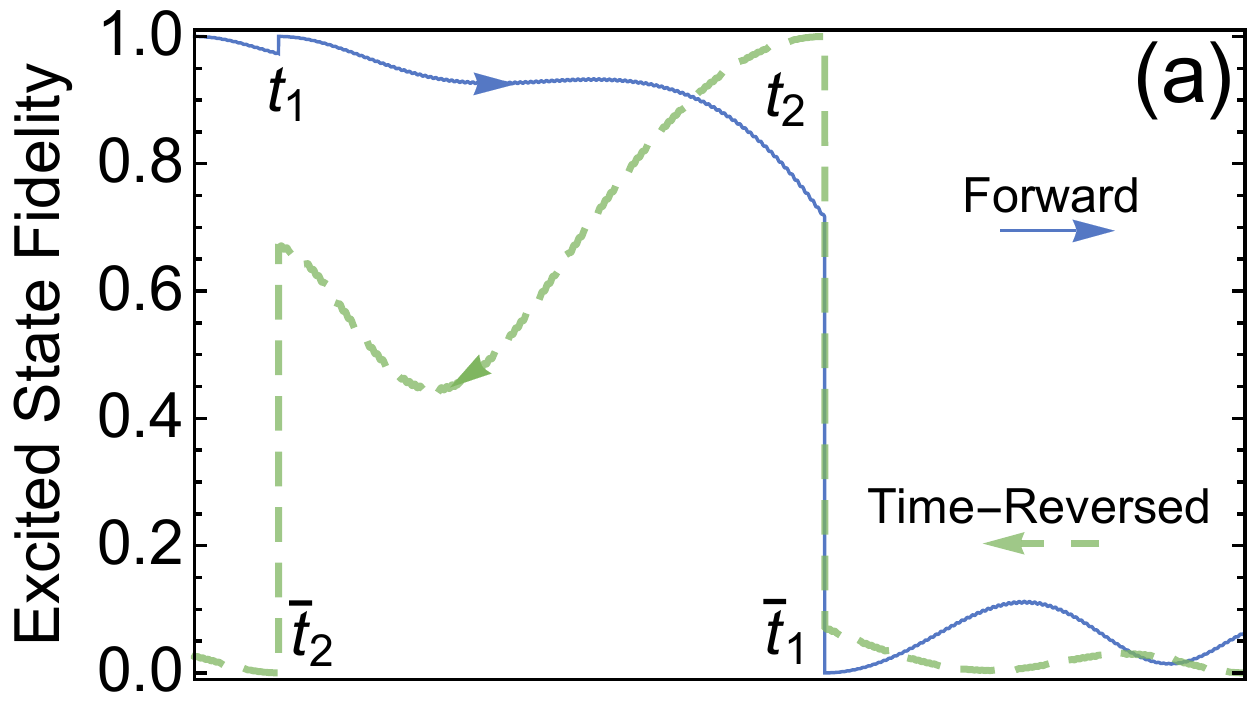}
        \includegraphics[height=2.62cm, clip]{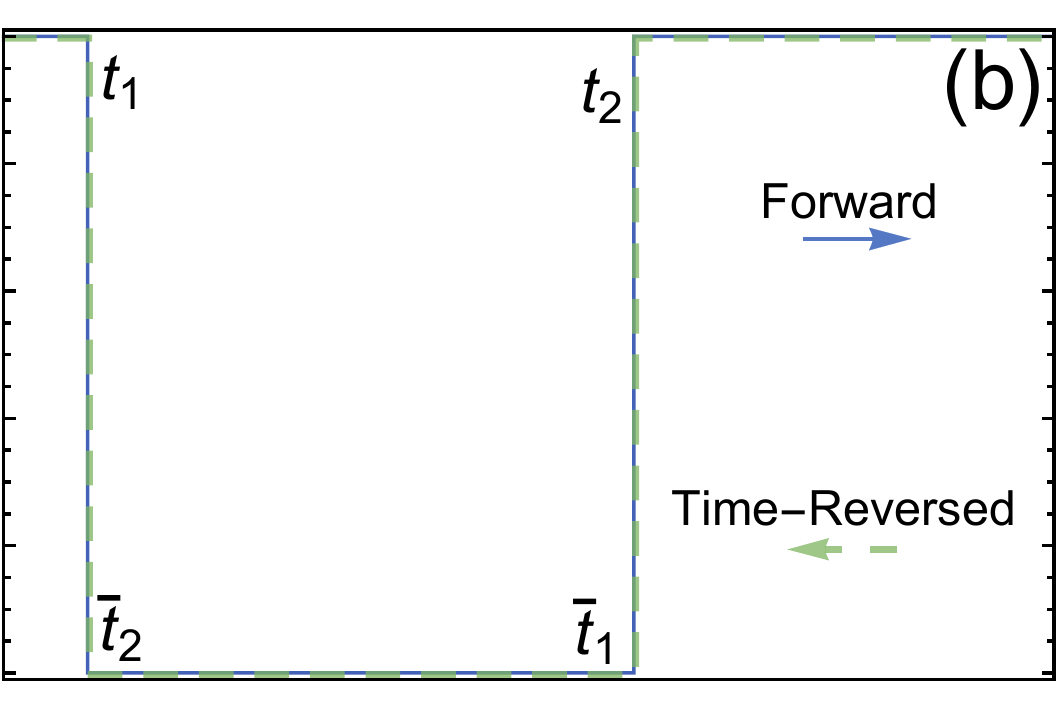}
      \end{center}
   \caption{(color online). (a) A quantum rare event in a spinless (time-reversal symmetric, i.e., $\Theta=I$) two-level system with $t_{\rm m}=0$ and $\rm M_A$ being the same as the initial PM. Here the outcome of the initial PM at time $t=0$ is assumed to be the excited state $e$. Nevertheless, an unexpected quantum jump occurs at $t_1$, indicated by a tiny jump in the excited state fidelity (blue). A large negative $I_{\rm QJT}$ is implied by the small excited state fidelity at the final state of the time-reversed QJT (green). (b) In the classical limit, the first jump at $t_1$ is always $e\to g$ if the system is initially in $e$ because of the absence of quantum superposition, and the time-reversed QJT is given by $|\bar\psi_t\rangle=|\psi_{\tau-t}\rangle$.}
   \label{fig3}
\end{figure}

The information content corresponding to Eq.~(\ref{GJE}) is found to be (see Appendix \ref{DIG} for the derivation)
\begin{equation}
I_{\rm{QJT}}[\psi_t,\alpha]=\ln\|\tilde M_\alpha|\bar\psi_{\bar t^-_{\mathrm{m}}}\rangle\|^2-\ln p_\alpha, 
\label{IG}
\end{equation}
which is the \emph{relevant information gain}, whose meaning will be explained latter. Here $|\bar\psi_t\rangle$ is the state at time $t$ in the TR QJT and uniquely determined by the forward QJT, and $p_\alpha=\mathrm{Tr}[M^\dag_\alpha M_\alpha\rho_{t^-_\mathrm{m}}]$ is the probability of the outcome $\alpha$ being observed for measurement $\mathrm{M_A}$. 
We note that for rank-$1$ measurements $I_{\rm QJT}$ can take on large negative values for \emph{quantum rare events}. For example, in a two-level system with states $e$ and $g$, the detection of the $g\to e$ jump can occur after a short time interval of \emph{coherent} driving conditioned on the initial energy projective outcome $e$ (see Fig.~\ref{fig3}). Such a rare event is a genuine quantum effect due to the fact that the system is brought into quantum superposition by coherent driving. Then the relevant information takes on a large negative value, reflecting our great surprise. Experimentally, $I_{\rm QJT}$ can be straightforwardly evaluated if we know the full details of the system. Otherwise, in principle it is still measurable, but in practice the measurement will be highly nontrivial (see Appendix \ref{DIG}).

Interestingly, when $\rm M_A$ is a unital channel (i.e., $\sum_\alpha M_\alpha M^\dag_\alpha=\sum_\alpha M^\dag_\alpha M_\alpha=I$), $\|\tilde M_\alpha|\bar\psi_{\bar t^-_{\mathrm{m}}}\rangle\|^2$ is the probability of the outcome $\alpha$ which is determined by the Bayesian inference based on the continuous monitoring results \emph{after} $t_{\rm m}$ in a single realization \cite{Molmer2013}, which is called \emph{retrodiction} \cite{Tsang2009} or retrofiltering \cite{Wiseman2015}. A bad retrodiction ensues from a quantum rare event. A simple interpretation for the emergence of the retrodiction probability rather than the usual prediction probability $\|M_\alpha|\psi_{t^-_{\rm m}}\rangle\|^2$ is that, retrodiction naturally encodes the effect of measurement backaction whereas prediction does not. 

The ensemble-averaged relevant information
\begin{equation}
\begin{split}
\langle I_{\rm{QJT}}\rangle=\sum_\alpha p_\alpha\mathcal{I}_{\mathrm{C}}(\rho^\alpha_{t^+_{\mathrm{m}}}:\Pi^{\lambda^\alpha_\tau}\mathrm{M}_{\mathrm{J}_{t_{\mathrm{m}}<t<\tau}|\alpha})\\
-\mathcal{I}_{\mathrm{C}}(\rho_{t^-_{\mathrm{m}}}:\Pi^{\lambda^\mathrm{A}_\tau} \mathrm{M}_{\mathrm{J}_{t_{\mathrm{m}}<t<\tau}|\mathrm{A}} \mathrm{M}_{\mathrm{A}})
\end{split}
\label{AIG}
\end{equation} 
gives a Holevo bound-like quantity (see Appendix \ref{DAIG} for details). Here $\mathcal{I}_{\mathrm{C}}(\rho:\mathrm{M_X})\equiv H(p^{\mathrm{M_X}}_\rho||p^{\mathrm{M_X}}_{\rho_\mathrm{u}})$, which we call the \emph{relevant information} of $\rho$ with respect to a general measurement $\mathrm{M_X}$ \cite{Wehrl1977,Balian1986,Szymusiak2016}, is the classical relative entropy \cite{Nielsen2010} between the $\mathrm{M_X}$ outcome probability distribution of $\rho$ (denoted by $p^{\mathrm{M_X}}_\rho$) and that of $\rho_{\mathrm{u}}\equiv I/d$; $\mathrm{M}_{\mathrm{J}_{t_{\mathrm{m}}<t<\tau}|\alpha}$ is the effective continuous measurement on the \emph{system} generated by $\mathcal{L}^\alpha_t$. Unlike the Shannon entropy of the outcomes (known as the Ingarden-Urbanik entropy \cite{Ingarden1962,Wehrl1978,Szymusiak2016}) which measures their \emph{uncertainty}, $\mathcal{I}_{\rm C}$ measures the extent to which we can \emph{specify the quantum state} based on the outcomes \cite{Balian1986}. Hence, $\langle I_{\rm QJT}\rangle$ measures the difference of our (average) knowledge on the selective post-measurement states $\rho^\alpha_{t^+_{\rm m}}=M_\alpha\rho_{t^-_{\rm m}}M^\dag_\alpha/p_\alpha$ and the pre-measurement state $\rho_{t^-_{\rm m}}$ acquired from \emph{all} the outcomes after $t^-_{\rm m}$. It is worth mentioning that $\mathcal{I}_{\mathrm{C}}$ was first mathematically introduced in Ref.~\cite{Wehrl1977}, and has enjoyed renewed interest recently in quantum information \cite{Szymusiak2016}. The applicability of $\mathcal{I}_{\mathrm{C}}$ to continuous measurements with $|\mathrm{X}|=\infty$ is based on the fact that $\mathcal{I}_{\mathrm{C}}(\rho:\mathrm{M_X})\le S(\rho||\rho_{\mathrm{u}})\equiv\mathcal{I}_{\mathrm{Q}}(\rho)$, where $S(\cdot||\cdot)$ is the quantum relative entropy \cite{Nielsen2010}.

Replacing all $\mathcal{I}_{\mathrm{C}}$ in Eq.~(\ref{AIG}) by $\mathcal{I}_{\mathrm{Q}}$, we obtain another upper bound of $-\beta\langle W_{\rm diss}\rangle$ called quantum-classical (QC)-mutual information $I_{\mathrm{QC}}$ \cite{Sagawa2008}, where $\langle W_{\mathrm{diss}}\rangle\equiv\langle W\rangle-\langle\Delta F\rangle$ is the dissipated work. While there is no magnitude relation between $I_{\mathrm{QC}}$ and $\langle I_{\rm{QJT}}\rangle$, as we will see in the next section, the latter (former) is expected to give a tighter (looser) bound, since it is (not) protocol-dependent and can be negative (is positive definite) if we carry out a bad feedback control. Nevertheless, the $I_{\mathrm{QC}}$ bound can be obtained from a fluctuation theorem for a \emph{different process} with the same $\langle W_{\rm diss}\rangle$  (see Appendix \ref{OFT}), and both $\langle I_{\rm{QJT}}\rangle$ and $I_{\mathrm{QC}}$ reproduce the same classical mutual information \cite{Sagawa2010} in the classical limit.

\section{Examples} 
\label{ex}
\subsection{Isolated two-level system}
We first consider a minimal model that demonstrates a quantum feedback control process: a pseudospin (so that $\Theta=I$) 
subjected to an effective magnetic field $\boldsymbol{B}=B(\cos\theta\boldsymbol{e}_z+\sin\theta\boldsymbol{e}_x)$ confined in the $x-z$ plane and isolated from any heat bath (adiabatic limit). The Hamiltonian of the system reads
\begin{equation}
H(\boldsymbol{B})=-\boldsymbol{\mu}\cdot\boldsymbol{B}=-\mu B(\cos\theta\sigma_z+\sin\theta\sigma_x).
\end{equation}
The initial state of the system is chosen to be the equilibrium state under the work parameter $\boldsymbol{B}_0=B_0\boldsymbol{e}_z$. After the initial PM, the system is purified as either $|\uparrow\rangle$ or $|\downarrow\rangle$, an eigenstate of $\sigma_z$, with probability $p^{\mathrm{eq}}_\uparrow=e^{\beta\mu B_0}/(2\cosh\beta\mu B_0)$ and $p^{\mathrm{eq}}_\downarrow=e^{-\beta\mu B_0}/(2\cosh\beta\mu B_0)$, respectively. Right after the initial PM, we perform a PM under the eigenbasis of $\sigma_z\cos\theta_0+\sigma_x\sin\theta_0$. If the outcome is $\uparrow_{\theta_0}$ ($\downarrow_{\theta_0}$), we quickly switch $\boldsymbol{B}_0$ to $\boldsymbol{B}_1=B_1(\cos\theta_1\boldsymbol{e}_z+\sin\theta_1\boldsymbol{e}_x)$ ($-\boldsymbol{B}_1$), immediately followed by the final PM. All the eight possible QJTs are listed in Table.~\ref{table1}. It is tedious but straightforward to check the validities of the two generalized quantum Jarzynski equalities (\ref{GJE}) analytically.

\begin{table*}[tbp]
\caption{Trajectory probability $\mathcal{P}[\psi_t,\alpha]$, probability of a measurement outcome $p_\alpha$, pre-measurement state in the time-reversed QJT $|\bar\psi_{\bar t^-_m}\rangle$, relevant information $I_{\rm QJT}[\psi_t,\alpha]$ and work $W[\psi_t,\alpha]$ along all the eight possible QJTs in the minimal model. Here
$p_{\uparrow_{\theta_0}}=p^{\mathrm{eq}}_\uparrow\cos^2{\frac{\theta_0}{2}}+p^{\mathrm{eq}}_\downarrow\sin^2{\frac{\theta_0}{2}}$ and
$p_{\downarrow_{\theta_0}}=p^{\mathrm{eq}}_\uparrow\sin^2{\frac{\theta_0}{2}}+p^{\mathrm{eq}}_\downarrow\cos^2{\frac{\theta_0}{2}}$ are respectively the probabilities to observe $\uparrow_{\theta_0}$ and $\downarrow_{\theta_0}$ when starting from $\rho_0=p^{\rm eq}_{\uparrow}|\uparrow\rangle\langle\uparrow|+p^{\rm eq}_{\downarrow}|\downarrow\rangle\langle\downarrow|$.}
\begin{center}
\begin{tabular}{ccccccccc}
\hline\hline
\;\;\;Initial\;\;\;& \;\;\;Feedback\;\;\; & \;\;\;Final\;\;\; & \;$\mathcal{P}[\psi_t,\alpha]$ \;& $\;\;p_\alpha\;\;$ & $\;\;|\bar\psi_{\bar t^-_{\rm m}}\rangle\;\;$ & $I_{\rm QJT}[\psi_t,\alpha]$ & $W[\psi_t,\alpha]$\\
\hline
\multirow{4}{*}{$\uparrow$} & \multirow{2}{*}{$\uparrow_{\theta_0}$} & $\uparrow_{\theta_1}$ & \;\;\;\;$p^{\mathrm{eq}}_\uparrow\cos^2\frac{\theta_0}{2}\cos^2\frac{\theta_0-\theta_1}{2}$\;\;\;\; & \multirow{2}{*}{\;\;\;\;$p_{\uparrow_{\theta_0}}
$\;\;\;\;} & \;\;\;\;$|\uparrow_{\theta_1}\rangle$\;\;\;\; & \;\;\;\;$\ln(\cos^2\frac{\theta_0-\theta_1}{2}/p_{\uparrow_{\theta_0}})$\;\;\;\; & \;\;$\mu(-B_1+B_0)$\;\;\\
& & $\downarrow_{\theta_1}$ & $p^{\mathrm{eq}}_\uparrow\cos^2\frac{\theta_0}{2}\sin^2\frac{\theta_0-\theta_1}{2}$  &  & $|\downarrow_{\theta_1}\rangle$ & $\ln(\sin^2\frac{\theta_0-\theta_1}{2}/p_{\uparrow_{\theta_0}})$ & $\mu(B_1+B_0)$\\
& \multirow{2}{*}{$\downarrow_{\theta_0}$} & $\uparrow_{\theta_1}$ & $p^{\mathrm{eq}}_\uparrow\sin^2\frac{\theta_0}{2}\cos^2\frac{\theta_0-\theta_1}{2}$ & \multirow{2}{*}{$p_{\downarrow_{\theta_0}}
$} & $|\downarrow_{\theta_1}\rangle$ & $\ln(\cos^2\frac{\theta_0-\theta_1}{2}/p_{\downarrow_{\theta_0}})$ & $\mu(-B_1+B_0)$\\
& & $\downarrow_{\theta_1}$ & $p^{\mathrm{eq}}_\uparrow\sin^2\frac{\theta_0}{2}\sin^2\frac{\theta_0-\theta_1}{2}$  &  & $|\uparrow_{\theta_1}\rangle$ & $\ln(\sin^2\frac{\theta_0-\theta_1}{2}/p_{\downarrow_{\theta_0}})$ & $\mu(B_1+B_0)$\\
\multirow{4}{*}{$\downarrow$} & \multirow{2}{*}{$\uparrow_{\theta_0}$} & $\uparrow_{\theta_1}$ & $p^{\mathrm{eq}}_\downarrow\sin^2\frac{\theta_0}{2}\cos^2\frac{\theta_0-\theta_1}{2}$ & \multirow{2}{*}{$p_{\uparrow_{\theta_0}}
$} & $|\uparrow_{\theta_1}\rangle$ & $\ln(\cos^2\frac{\theta_0-\theta_1}{2}/p_{\uparrow_{\theta_0}})$ & $\mu(-B_1-B_0)$\\
& & $\downarrow_{\theta_1}$ & $p^{\mathrm{eq}}_\downarrow\sin^2\frac{\theta_0}{2}\sin^2\frac{\theta_0-\theta_1}{2}$  & & $|\downarrow_{\theta_1}\rangle$ & $\ln(\sin^2\frac{\theta_0-\theta_1}{2}/p_{\uparrow_{\theta_0}})$ & $\mu(B_1-B_0)$\\
& \multirow{2}{*}{$\downarrow_{\theta_0}$} & $\uparrow_{\theta_1}$ & $p^{\mathrm{eq}}_\downarrow\cos^2\frac{\theta_0}{2}\cos^2\frac{\theta_0-\theta_1}{2}$ & \multirow{2}{*}{$p_{\downarrow_{\theta_0}}
$} & $|\downarrow_{\theta_1}\rangle$ & $\ln(\cos^2\frac{\theta_0-\theta_1}{2}/p_{\downarrow_{\theta_0}})$ & $\mu(-B_1-B_0)$\\
& & $\downarrow_{\theta_1}$ & $p^{\mathrm{eq}}_\downarrow\cos^2\frac{\theta_0}{2}\sin^2\frac{\theta_0-\theta_1}{2}$  &  & $|\uparrow_{\theta_1}\rangle$ & $\ln(\sin^2\frac{\theta_0-\theta_1}{2}/p_{\downarrow_{\theta_0}})$ & $\mu(B_1-B_0)$\\
\hline\hline
\end{tabular}
\end{center}
\label{table1}
\end{table*}

\begin{figure}
\begin{center}
        \includegraphics[width=4.2cm, clip]{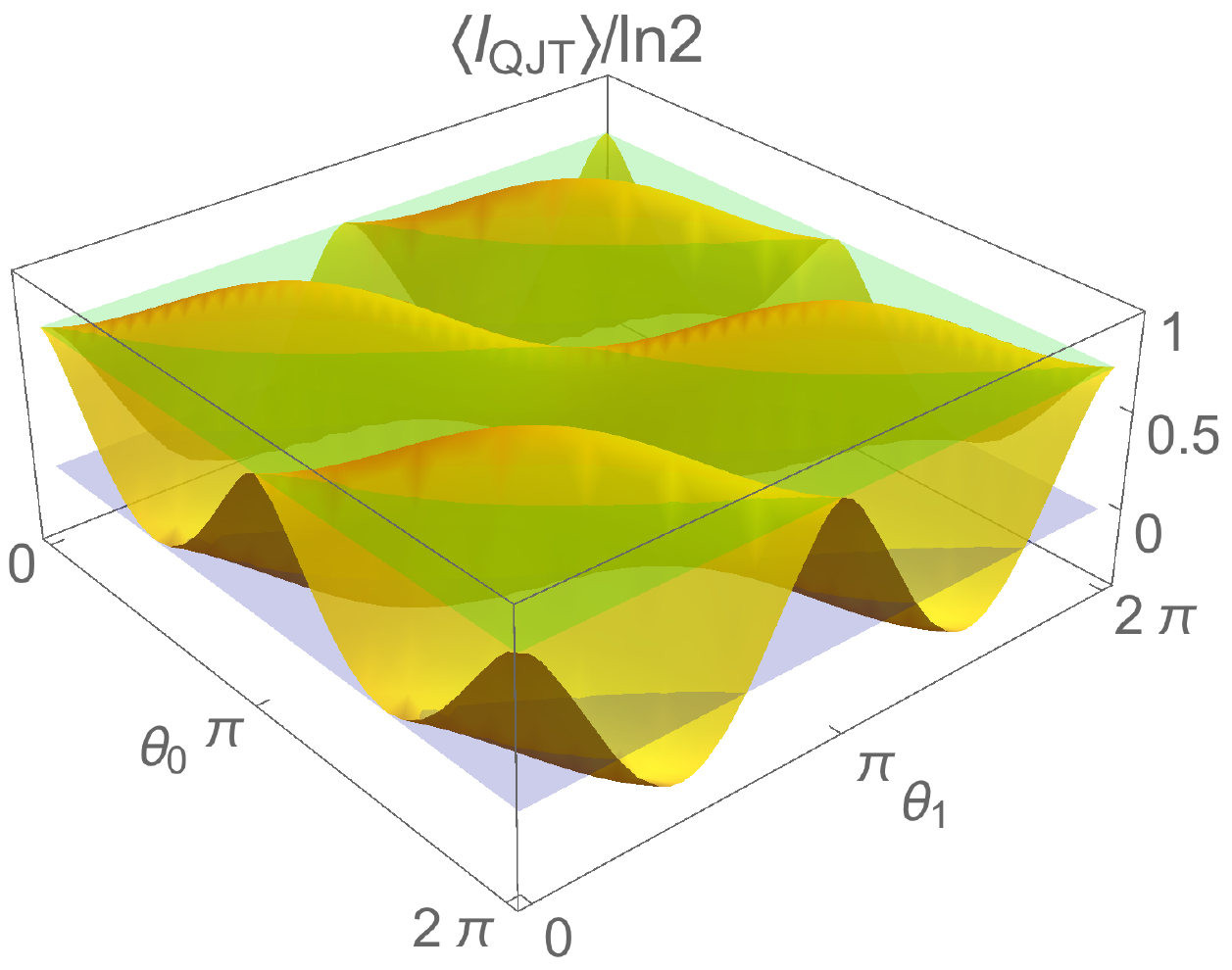}
        \includegraphics[width=4.2cm, clip]{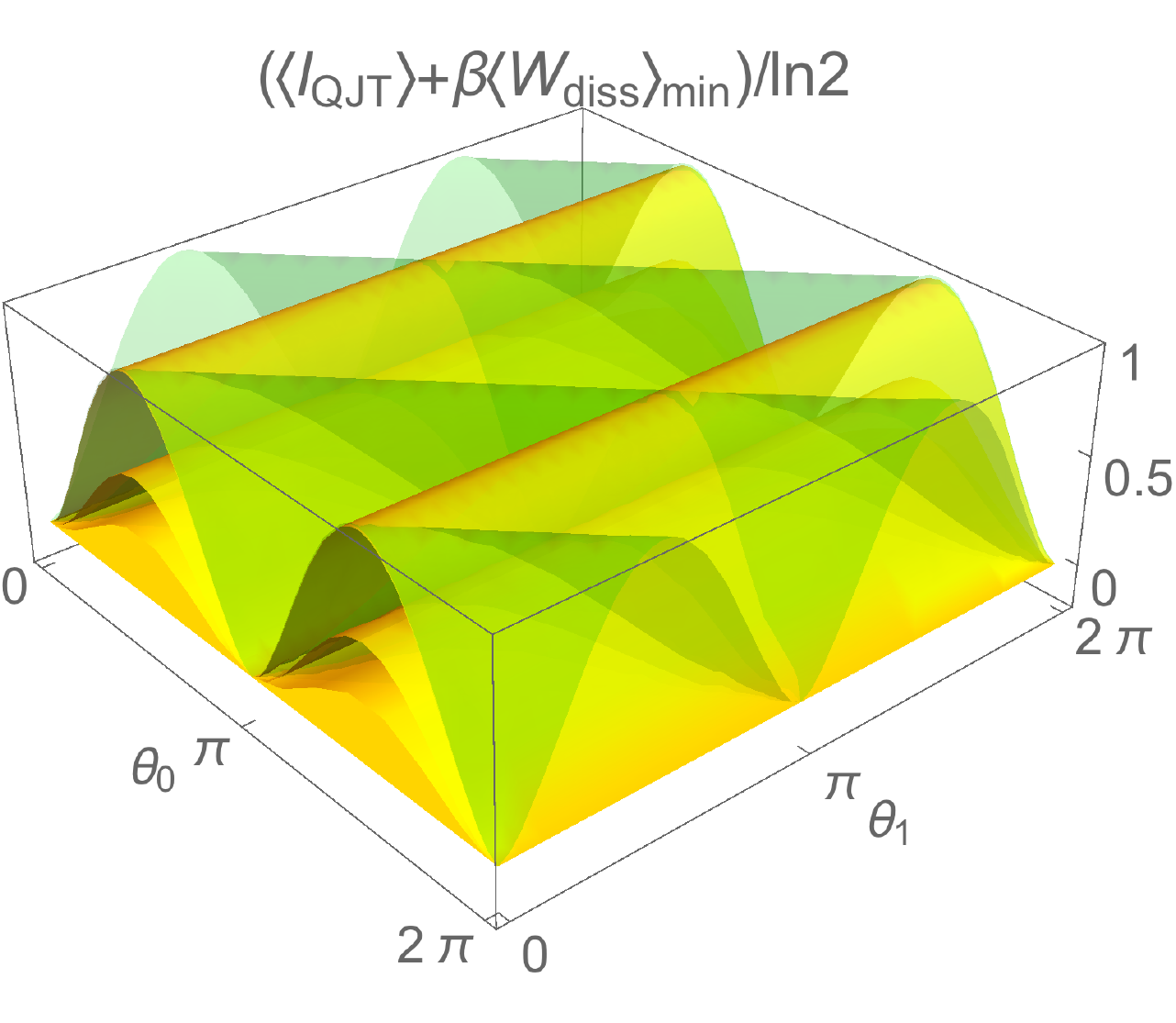} 
      \end{center}
   \caption{$\langle I_{\rm QJT}\rangle/\ln 2$ (left) and $(\langle I_{\rm QJT}\rangle+\langle W_{\rm diss}\rangle_{\rm{max}})/\ln 2$ (right) in the $\theta_0$-$\theta_1$ parameter space. In the left figure, the green and blue plane respectively correspond to the QC-mutual information $I_{\rm QC}$ and 0, $p^{\rm eq}_{\uparrow}$ is fixed as $0.8$. In the right figure, the gree curved surface refers to $I_{\rm QC}$ (overestimation from the exact $-\beta\langle W_{\rm diss}\rangle_{\rm min}$), while the remaining yellow ones show $\langle I_{\rm QJT}\rangle$ for different equilibrium initial states ($p^{\rm eq}_{\uparrow}=0.5,0.8,0.9,0.999$ from the lowest to the highest).}
\label{fig4}
\end{figure}

After a few analytical calculations, we obtain the following expressions of $\langle I_{\rm QJT}\rangle$ and $I_{\rm QC}$
\begin{equation}
\begin{split}
\langle I_{\rm QJT}\rangle&=H_2\left(p_{\uparrow_{\theta_0}}\right)-H_2\left(\cos^2\frac{\theta_0-\theta_1}{2}\right),\\
I_{\rm QC}&=H_2\left(p^{\mathrm{eq}}_\uparrow\right),
\end{split}
\end{equation}
where $p_{\uparrow_{\theta_0}}=p^{\mathrm{eq}}_\uparrow\cos^2\frac{\theta_0}{2}+p^{\mathrm{eq}}_\downarrow\sin^2\frac{\theta_0}{2}$ and $H_2(x)\equiv -x\ln x-(1-x)\ln(1-x)$. For a special case with $p^{\mathrm{eq}}_\uparrow=0.8$, we draw the curved surface of $\langle I_{\rm QJT}\rangle$ with respect to $\theta_{1,2}$ in Fig.~\ref{fig4} (left), which turns out to be larger than $I_{\rm QC}$ (less than $0$) in some regions. Thus, there is no general magnitude relation between $\langle I_{\rm QJT}\rangle$ and $I_{\rm QC}$ ($0$).

Besides the absence of a universal magnitude relation, the model also shows that the upper bound $\beta^{-1}\langle I_{\rm QJT}\rangle$ for the minus dissipated work $-\langle W_{\rm diss}\rangle\equiv-\langle W\rangle +\langle\Delta F\rangle$ is not globally achievable (unless $p^{\rm eq}_\uparrow=0.5$). By minimizing $\langle W\rangle$ (maximizing $-\langle W\rangle$) for a given $p^{\mathrm{eq}}_\uparrow$, we obtain
\begin{equation}
-\beta\langle W_{\mathrm{diss}}\rangle_{\mathrm{min}}=H_2\left(p^{\mathrm{eq}}_\uparrow\right)-H_2\left(\cos^2\frac{\theta_0-\theta_1}{2}\right).
\label{exactbound}
\end{equation}
For $p^{\mathrm{eq}}_\uparrow=0.5,0.8,0.9,0.999$, we draw the difference (subtraction) between the bound given by $\langle I_{\rm QJT}\rangle$ (or $\langle I_{\rm QC}\rangle$) and the exact $-\beta\langle W_{\rm diss}\rangle_{\rm min}$ (\ref{exactbound}) in Fig.~\ref{fig4} (right). For $p^{\rm eq}_\downarrow=0.5$, $\langle I_{\rm QJT}\rangle$ coincides with the exact bound (lowest plane). As the initial entropy decreases, the estimation of $\langle I_{\rm QJT}\rangle$ becomes worse, while $I_{\rm QC}+\beta\langle W_{\rm diss}\rangle_{\rm min}=H_2(\cos^2[(\theta_0-\theta_1)/2])$ is independent of $p^{\rm{eq}}_\uparrow$ (green curved surface). Generally speaking, $\langle I_{\rm QJT}\rangle$ is a better bound than $I_{\rm QC}$, since it involves the information of the concrete feedback control protocols.

\subsection{Dissipative two-level system}

\begin{figure}
\begin{center}
        \includegraphics[width=8cm, clip]{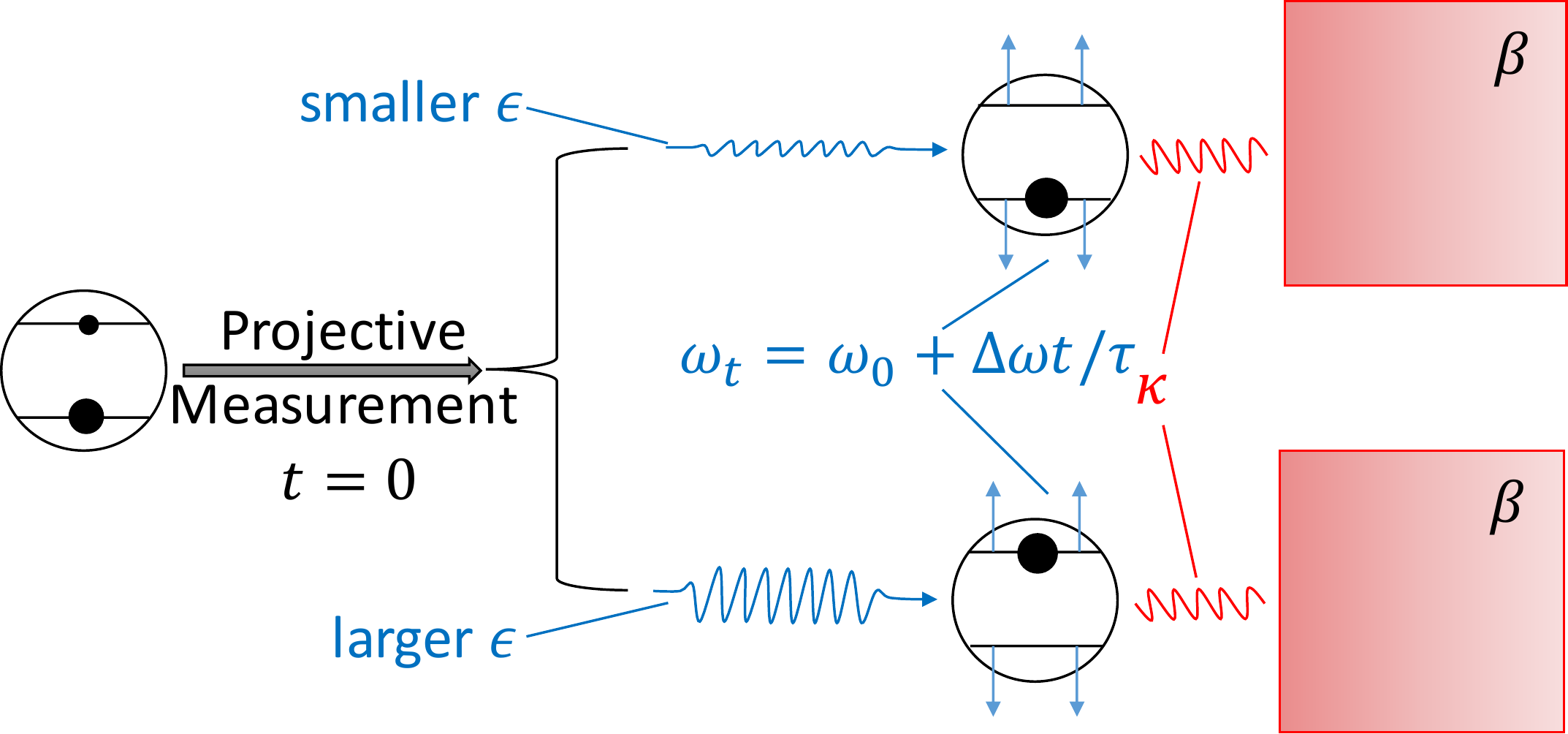}\\
      \end{center}
   \caption{(color online). Feedback control of a dissipative two-level system based on the initial projective measurement. The strength of the coherent driving $\epsilon$ is tuned to be smaller (larger) if the outcome is the ground state (excited state). The inclusive driving protocol $\omega_t=\omega_0+\Delta\omega t/\tau$ and the coupling strength to the heat bath are the same.}
   \label{fig5}
\end{figure}

Since our findings are generally applicable to open quantum systems, let us consider a dissipative two-level system under coherent driving, where the equation of motion reads (see Appendix \ref{EOME})
\begin{equation}
\dot\rho_t=-\frac{i}{2}[\omega_t\sigma_z+\epsilon\sigma_x\cos\omega_{\mathrm{d}} t,\rho_t]+\sum_{j=\pm}\gamma_j(\omega_t)\mathcal{D}[\sigma_j]\rho_t.
\label{TLSEOM}
\end{equation}
Here the unitary part consists of the inclusive Hamiltonian $H(\omega)=\hbar\omega\sigma_z/2$ and the exclusive driving $h_t=\epsilon\sigma_x\cos\omega_{\mathrm{d}} t/2\ll H(\omega)$, $\sigma_\pm\equiv(\sigma_x\pm i\sigma_y)/2$ is the excitation (de-excitation) jump operator, and the corresponding transition rate $\gamma_\pm(\omega)=\kappa\omega[\coth(\beta\hbar\omega/2)\mp1]/2$ ensures the detailed balance condition. To perform feedback control, we perform the initial error-free PM, and then apply a weaker (stronger) external perturbation if the outcome is the ground (excited) state (see Fig.~\ref{fig5}). In this way, we can suppress (enhance) the probability of no jump events from the initial ground (excited) state to the final excited (ground) state. These events greatly contribute positive (negative) work values. Here, we choose a linear protocol $\omega_t=\omega_0+\Delta\omega t/\tau$ and a driving frequency $\omega_{\mathrm{d}}=0.1\pi$ with $\omega_0=0.3,\;\Delta\omega=0.1$, and $\tau=2000$. The driving strength is tuned to be $\epsilon=0.002$ ($0.008$)$\ll\omega_0$ for the ground (excited) initial state. The inverse temperature and the coupling strength are fixed at $\beta=5$ and $\kappa=0.001$, respectively.

\begin{figure}
\begin{center}
        \includegraphics[height=3.2cm, clip]{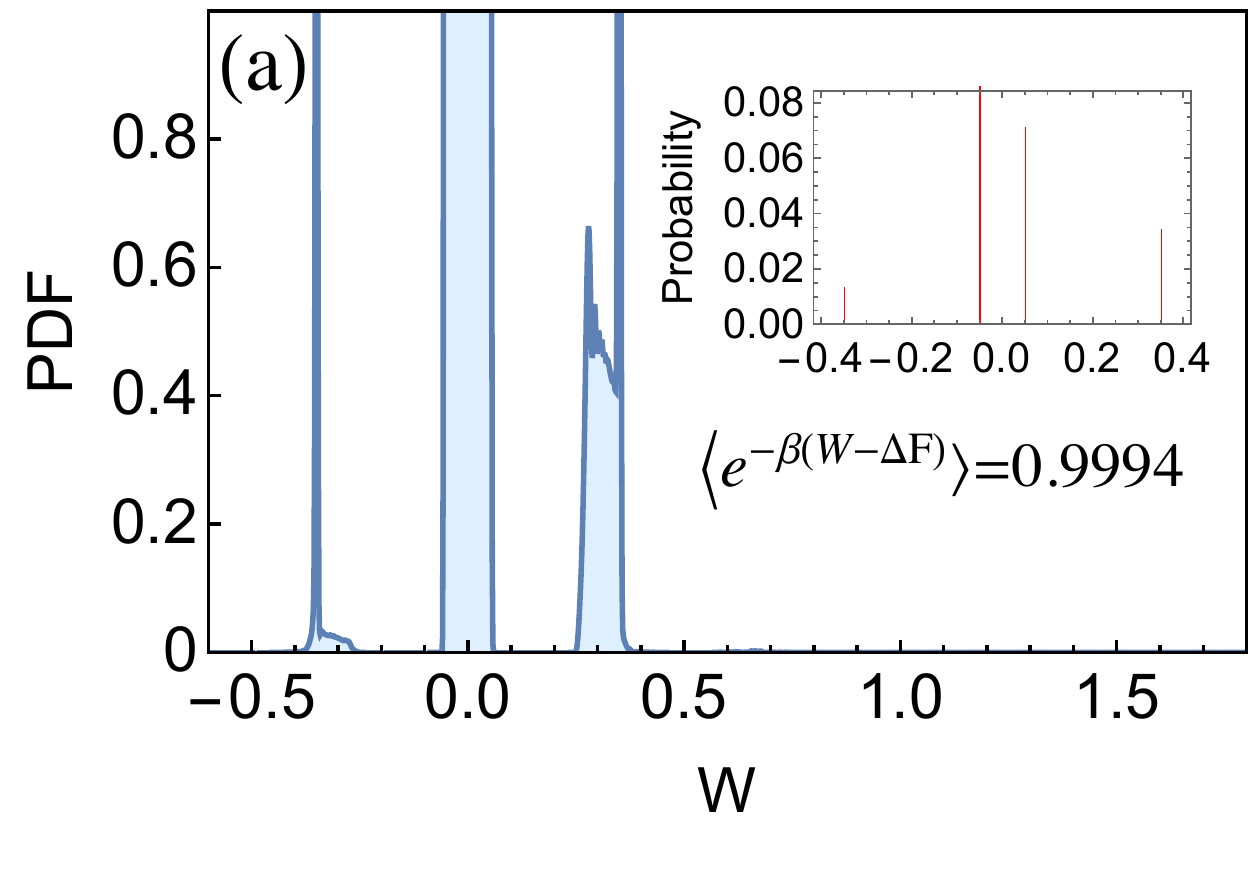}
        \includegraphics[height=3.2cm, clip]{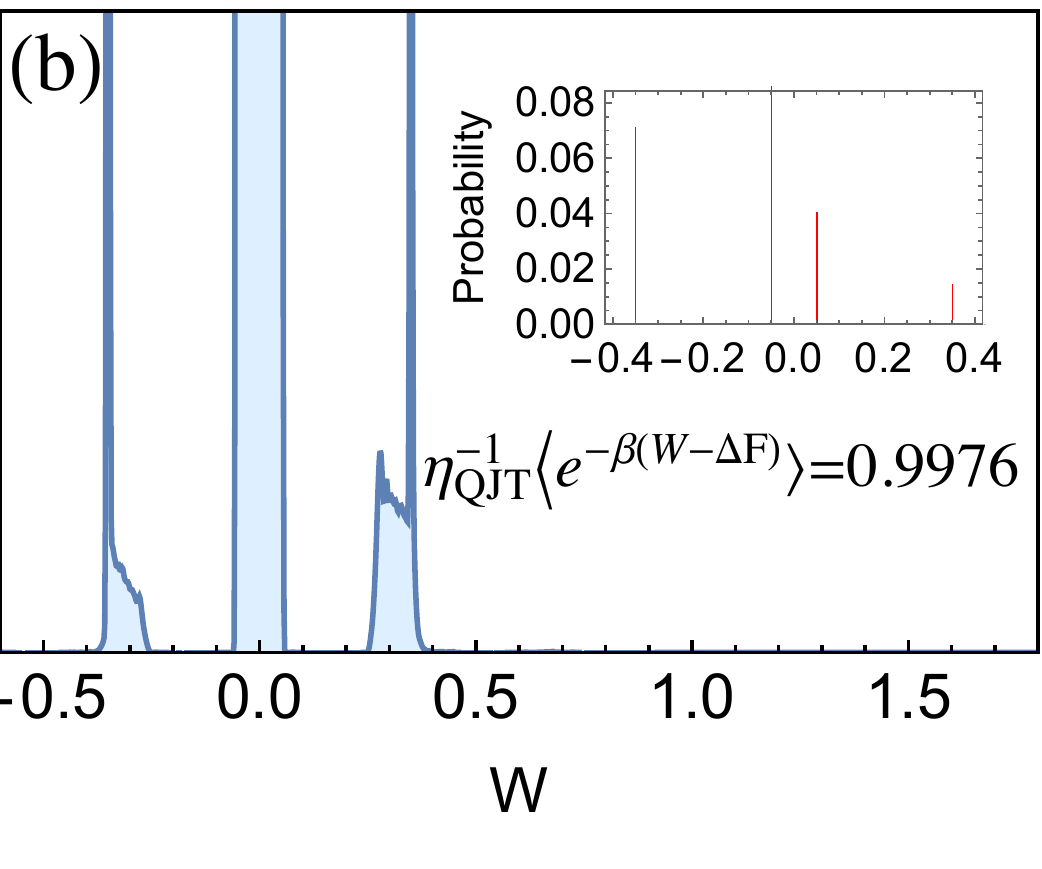} \ \
        \includegraphics[height=3.2cm, clip]{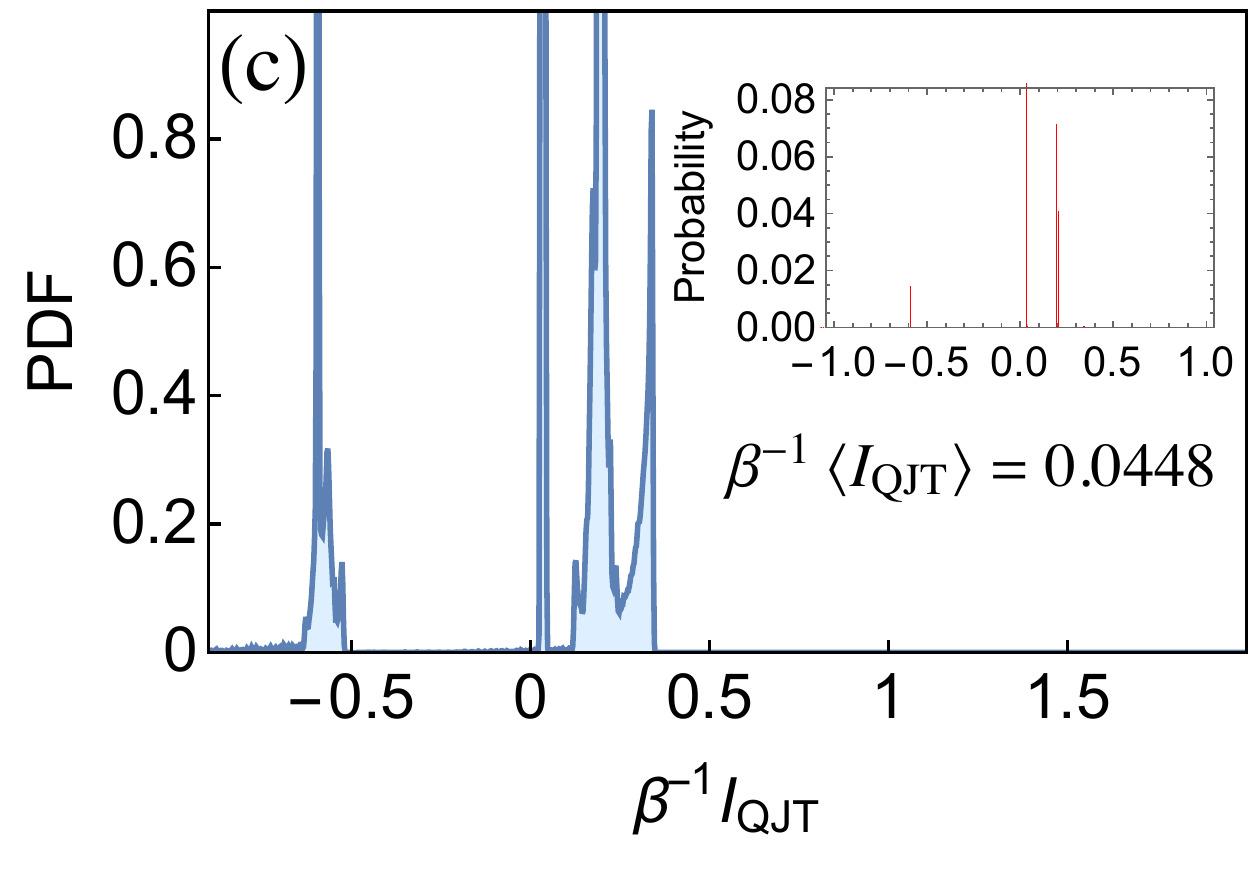}
        \includegraphics[height=3.2cm, clip]{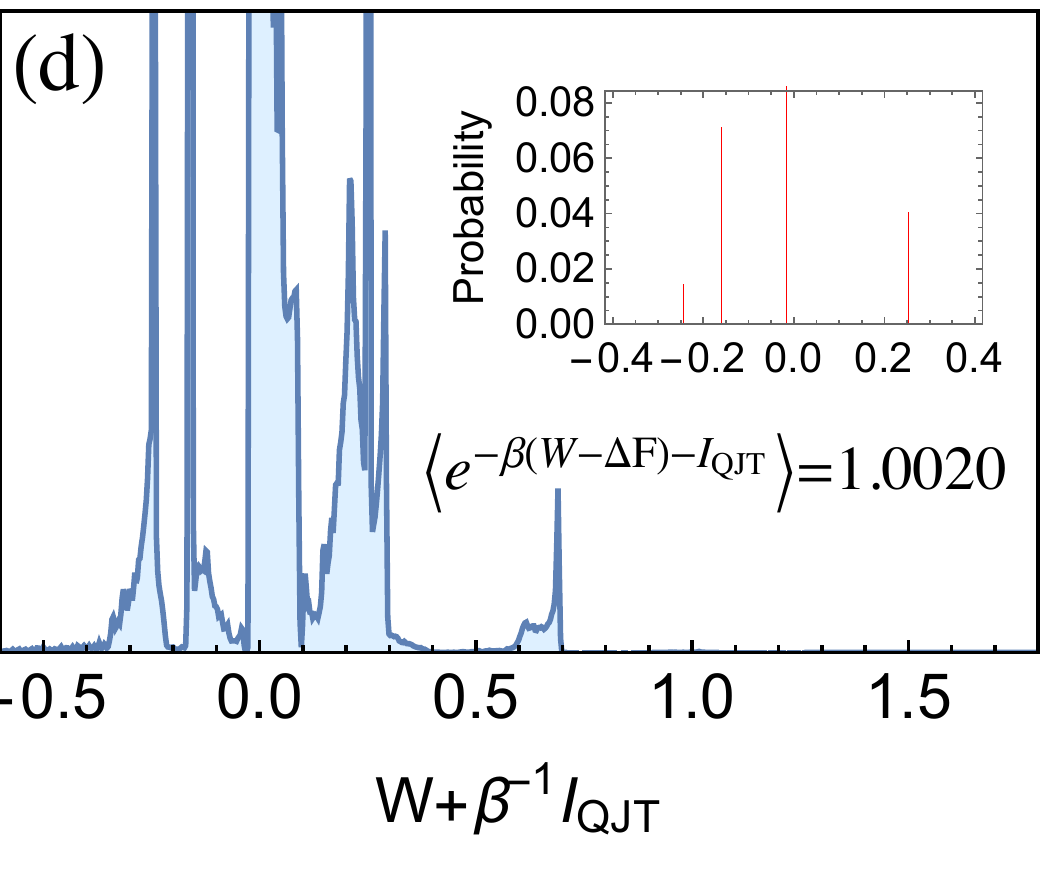}
      \end{center}
   \caption{(color online). Numerical verification of the generalized quantum Jarzynski equalities in a dissipative two-level system under coherent driving (\ref{TLSEOM}). Work distributions (a) without and (b) with feedback control. The distributions of (c) $I_{\rm QJT}$ (\ref{IG}), and (d) the composite variable $W+\beta^{-1}I_{\rm QJT}$ for the feedback control process. Insets show the probabilities of the divergent $\delta$-type peaks.}
   \label{fig6}
\end{figure}

We numerically evaluate (see Appendix \ref{NC}) the probability density functions (PDFs) of work, $\beta^{-1}I_{\rm QJT}$ and their sum as shown in Fig.~\ref{fig6} (b)-(d). For comparison, the work statistics of the corresponding ordinary driving process, with the same protocol $\omega_t$ but a fixed $\epsilon=0.0031$, is shown in Fig.~\ref{fig6} (a). Qualitatively, we observe both continuous parts (described by the probability density) and $\delta$-type peaks (described by the probability) in the work distributions, including the $\delta$-peaks caused by coherent driving, showing a combined nature of the work statistics in classical and isolated quantum systems. Comparing Fig.~\ref{fig6} (b) with Fig.~\ref{fig6} (a), we find that the rightmost (leftmost) $\delta$-type peak, corresponding to the QJTs connecting the initial ground (excited) state to the final excited (ground) state with no jumps, is considerably suppressed (enhanced). Quantitatively, we verify Eq.~(\ref{GJE}) with reasonable accuracy. At the ensemble level, the mean dissipated work $\langle W_{\mathrm{diss}}\rangle=-0.0139$ ($0.0244$) for the feedback control (ordinary) process, implying an apparent violation (the validity) of the second law. On the other hand, $-\langle W_{\mathrm{diss}}\rangle$ is far from saturating the upper bound $\beta^{-1}\langle I_{\rm{QJT}}\rangle=0.0448$ (much tighter than $\beta^{-1}I_{\rm QC}=0.0950$), indicating that the process is highly nonequilibrium.

In fact, we have chosen the parameters which are experimentally accessible in a superconducting qubit system \cite{Schon2001} such as a Cooper-pair box with a SQUID geometry, where $\omega$ can be tuned by varying the gate voltage, while the coherent driving is achievable by a rapidly oscillating magnetic flux through the SQUID \cite{Hakonen2010}. 
Superconducting qubits operate in a highly controllable way, especially a measurement can be performed very fast. Also, quantum jumps have been observed via coupling to a readout device \cite{Siddiqi2011}, which may simultaneously serve as an effective heat bath \cite{Siddiqi2012}. Therefore, despite the fact that measuring quantum work and heat statistics are still challenging \cite{Paternostro2014,Quan2015}, superconducting qubit systems should provide an ideal playground to investigate quantum information thermodynamics at the trajectory level. We note that there is an experimental proposal to study the energy fluctuations in a superconducting qubit, where only the technique of PM is required \cite{Pekola2015a}.

\section{Conclusions} 
\label{conclu}
We have developed a general framework to study the thermodynamics of open quantum systems with discrete feedback control at the level of individual QJTs. In particular, we have derived the generalized quantum Jarzynski equalities, which qualitatively differ from the classical counterparts due to quantum coherence and measurement backaction. We have proposed a minimal model of a two-level isolated system to analyze the performance of the new information content compared with the QC-mutual information. We have also numerically computed explicit work distributions in a dissipative two-level system driven out of equilibrium as a simple, nontrivial and experimentally accessible model, to verify the derived fluctuation theorems.

\begin{acknowledgements}
We acknowledge K. M{\o}lmer for valuable discussions. This work was supported by KAKENHI Grant No. 26287088 from the Japan Society for the Promotion of Science, a Grant-in-Aid for Scientific Research on Innovative Areas ``Topological Materials Science" (KAKENHI Grant No. 15H05855), the Photon Frontier Network Program from MEXT of Japan, and ImPACT Program of Council for Science, Technology and Innovation (Cabinet Office, Government of Japan). Z.G. was supported by MEXT scholarship. Y.A. was supported by the Japan Society for the Promotion of Science through Program for Leading Graduate Schools (ALPS). 
\end{acknowledgements}

\appendix
\section{Explicit expression of the Lindbald master equation (\ref{EOM})}
\label{apdA}
The fundamental equation of motion (\ref{EOM}) in the main text is a mixture of the perturbative Lindblad master equation (LME) and the adiabatic LME. The conventional perturbative LME is obtained if we simply add the perturbative term into the unitary part of a LME with a \emph{time-independent} generator, which is reasonable as long as the system-heat bath interaction is almost unaffected by the small (and usually rapidly oscillating) perturbation \cite{Carmichael2002,Silaev2014,Liu2014a}. Owing to the same argument, this straightforward modification should also be applicable to the cases with instantaneous disturbance (no longer perturbative) and/or slow variations of the work parameter in the adiabatic regime. Therefore, under appropriate conditions, we can staightforwardly write down the explicit expression of Eq.~(\ref{EOM}) for a given $h_t$ once we know the underlying adiabatic LME. We emphasize that this simple modification cannot be applied to the cases with strong driving fields ($h_t\sim H(\lambda_t)$) \cite{Alicki2013,Esposito2015}.

A detailed derivation of a general adiabatic LME \emph{starting from the Schr\"odinger equation alone} is given in Ref.~\cite{Lidar2012}. Here we just present the main result and show how it can be transformed into Eq.~(\ref{EOM}) in the main text. Let us consider a general ``small system + large environment" Hamiltonian:
\begin{equation}
H_{\mathrm{tot}}(t)=H_{\mathrm{S}}(\lambda_t)\otimes I_{\mathrm{B}}+I_{\mathrm{S}}\otimes H_{\mathrm{B}}+g\sum_{\alpha} A_\alpha\otimes B_\alpha,
\end{equation}
where $H_{\mathrm{S}}(\lambda_t)$ and $H_{\mathrm{B}}$ are respectively the bare Hamiltonians of the system and the heat bath, $A_\alpha$ and $B_\alpha$ are all dimensionless Hermitian operators and $g$ is the coupling strength with the dimension of energy. The typical energy gap of $H_{\mathrm{S}}(\lambda)$ is denoted by $\Delta(\lambda)$, and the typical decay time of the correlation function of the heat bath $\mathcal{B}_{\alpha\beta}(t)\equiv\mathrm{Tr}[B_\alpha(t)B_\beta(0)\rho^{\mathrm{eq}}_{\mathrm{B}}]$, namely the memory time of the heat bath, is denoted by $\tau_{\mathrm{B}}$, where we introduce $B_\alpha(t)\equiv e^{iH_{\mathrm{B}}t/\hbar}B_\alpha e^{-iH_{\mathrm{B}}t/\hbar}$ and $\rho^{\mathrm{eq}}_{\mathrm{B}}\equiv e^{-\beta H_{\mathrm{B}}}/Z_{\mathrm{B}}$ with $Z_{\mathrm{B}}\equiv\mathrm{Tr}[e^{-\beta H_{\mathrm{B}}}]$. Defining $q(\lambda)\equiv\max_{a\neq b}|\langle a^\lambda|\partial_\lambda H_{\mathrm{S}}(\lambda)|b^\lambda\rangle|$ ($|a^\lambda\rangle$ or $|b^\lambda\rangle$ is an eigenstate of $H(\lambda)$), we impose the following conditions for the whole process for $t\in[0,\tau]$:
\begin{equation}
\begin{split}
\frac{\tau_{\mathrm{B}}}{\Delta(\lambda_t)}q(\lambda_t)\dot\lambda_t&\ll\min\left\{\frac{\Delta(\lambda_t)\tau_{\mathrm{B}}}{\hbar},\frac{\hbar}{\Delta(\lambda_t)\tau_{\mathrm{B}}}\right\},\\
\frac{g\tau_{\mathrm{B}}}{\hbar}&\ll\min\left\{1,\frac{\Delta(\lambda_t)}{g}\right\},
\end{split}
\label{judgement}
\end{equation}
which provide an appropriate separation of time scales. Under such conditions, after the standard Born-Markov \footnote{The heat bath is assumed to be always at equilibrium in the Born approximation, i.e., the total density operator is assumed to be $\rho_t\otimes\rho^{\rm eq}_{\rm B}$, $\rho^{\rm eq}_{\rm B}\equiv e^{-\beta H_{\rm B}}/Z_{\rm B}$ during the whole process. This is the origin of the detailed balance condition.} and the rotating-wave approximations, the following adiabatic LME can be derived:
\begin{widetext}
\begin{equation}
\begin{split}
\dot\rho_t=-\frac{i}{\hbar}[H_{\mathrm{S}}(\lambda_t)+H_{\mathrm{LS}}(\lambda_t),\rho_t]&+\sum_{\alpha,\beta,a\neq b}\gamma_{\alpha\beta}(\omega^{\lambda_t}_{ba})\left[L_{ab,\beta}(\lambda_t)\rho_tL^\dag_{ab,\alpha}(\lambda_t)-\frac{1}{2}\{L^\dag_{ab,\alpha}(\lambda_t)L_{ab,\beta}(\lambda_t),\rho_t\}\right]\\
&+\sum_{\alpha,\beta,a,b}\gamma_{\alpha\beta}(0)[L_{aa,\beta}(\lambda_t)\rho_tL^\dag_{bb,\alpha}(\lambda_t)-\frac{1}{2}\{L^\dag_{aa,\alpha}(\lambda_t)L_{bb,\beta}(\lambda_t),\rho_t\}],
\end{split}
\label{ALME}
\end{equation}
\end{widetext}
where $\hbar\omega^\lambda_{ba}=E^\lambda_b-E^\lambda_a$ is the energy difference between the $b$-th and the $a$-th energy levels of the system, $L_{ab,\alpha}(\lambda)=A_{ab,\alpha}(\lambda)|a^\lambda\rangle\langle b^\lambda|$ with $A_{ab,\alpha}(\lambda)\equiv\langle a^\lambda|A_\alpha|b^\lambda\rangle$, $\gamma_{\alpha\beta}(\omega)=\frac{g^2}{\hbar^2}\int^\infty_{-\infty}dt e^{i\omega t}\mathcal{B}_{\alpha\beta}(t)$ is Hermitian and satisfies the detailed balance condition $\gamma_{\alpha\beta}(-\omega)=e^{-\beta\hbar\omega}\gamma_{\beta\alpha}(\omega)$, and $H_{\mathrm{LS}}(\lambda)=\sum E^b_{\mathrm{LS}}(\lambda) |b^\lambda\rangle\langle b^\lambda|$ describes the Lamb shift Hamiltonian reads, where
\begin{equation}
\begin{split}
&E^b_{\mathrm{LS}}(\lambda)=\sum_{\alpha,\beta,a}A^*_{ab,\alpha}(\lambda) S_{\alpha\beta}(\omega^\lambda_{ba})A_{ab,\beta}(\lambda),\\
&S_{\alpha\beta}(\omega)=\int^\infty_{-\infty}\frac{d\omega'}{2\pi}\mathcal{P}\frac{\gamma_{\alpha\beta}(\omega')}{\omega-\omega'},
\end{split}
\end{equation}
with $\cal{P}$ denoting the principal value. The first summation in Eq.~(\ref{ALME}) can be simplified as 
\begin{equation}
\begin{split}
&\sum_{a\neq b} \left[L_{ab}(\lambda)\rho_tL^\dag_{ab}(\lambda)-\frac{1}{2}\{L^\dag_{ab}(\lambda)L_{ab}(\lambda),\rho_t\}\right]\\
&=\sum_{a\neq b}\mathcal{D}[L_{ab}(\lambda)]\rho_t, 
\end{split}
\end{equation}
where
\begin{equation}
\begin{split}
&L_{ab}(\lambda)=\sqrt{w_{ba}(\lambda)}|a^\lambda\rangle\langle b^\lambda|,\\
&w_{ba}(\lambda)=\sum_{\alpha,\beta} A^*_{ab,\alpha}(\lambda)\gamma_{\alpha\beta}(\omega^\lambda_{ba})A_{ab,\beta}(\lambda).
\end{split}
\label{Lab}
\end{equation}
The jump operators $L_{ab}(\lambda)$, satisfying $[L_{ab}(\lambda),H_{\mathrm{S}}(\lambda)]=\hbar\omega^\lambda_{ba}L_{ab}(\lambda)$, $L^\dag_{ba}(\lambda)=L_{ab}(\lambda)e^{-\beta\hbar\omega^\lambda_{ba}/2}$ \footnote{Notice that $w_{ba}(\lambda)=\sum_{\alpha,\beta} A^*_{ba,\alpha}(\lambda)\gamma_{\alpha\beta}(\omega^\lambda_{ab})A_{ba,\beta}(\lambda)=\sum_{\alpha,\beta} A_{ab,\alpha}(\lambda)\gamma_{\beta\alpha}(\omega^\lambda_{ba})e^{-\beta\hbar\omega^\lambda_{ba}}A^*_{ab,\beta}(\lambda)=w_{ba}(\lambda)e^{-\beta\hbar\omega^\lambda_{ba}}$, and $L_{ab}(\lambda)=\sqrt{w_{ba}(\lambda)}|a^\lambda\rangle\langle b^\lambda|$.}, are related to dissipation (i.e., nonzero energy exchange with the heat bath), where $w_{ba}(\lambda)$ is real and positive, which can be interpreted as the transition rate from the $b$-th eigenstate to the $a$-th one. The second sum in Eq.~(\ref{ALME}) can be simplified as 
\begin{equation}
\begin{split}
&\sum_{\sigma} \left[L_\sigma(\lambda)\rho_tL^\dag_\sigma(\lambda)-\frac{1}{2}\{L^\dag_\sigma(\lambda)L_\sigma(\lambda),\rho_t\}\right]\\
&=\sum_\sigma\mathcal{D}[L_\sigma(\lambda)]\rho_t,
\end{split}
\end{equation}
where
\begin{equation}
\begin{split}
&L_\sigma(\lambda)=\sqrt{\gamma_\sigma(0)}\sum_{a,\alpha} o_{\sigma\alpha} A_{aa,\alpha}(\lambda)|a^\lambda\rangle\langle a^\lambda|,\\
&\gamma_{\alpha\beta}(0)=\sum_\sigma o_{\sigma\alpha}o_{\sigma\beta}\gamma_\sigma(0),
\end{split}
\end{equation}
with $o_{\alpha\beta}$'s being the elements of the orthogonal matrix that diagonalizes the real symmetric and positive definite matrix $\gamma_{\alpha\beta}(0)$ \footnote{This is due to the Hermitian property of $\gamma_{\alpha\beta}(\omega)$, i.e., $\gamma^*_{\beta\alpha}(\omega)=\gamma_{\alpha\beta}(\omega)$, as well as the detailed balance condition $\gamma_{\alpha\beta}(-\omega)=\gamma_{\beta\alpha}(\omega)e^{-\beta\hbar\omega}$. $\gamma_{\alpha\beta}(\omega)$ is also positive definite, which ensures $w_{ba}(\lambda)$ to be positive, and can be proved by using the Lehmann representation.}. These jump operators, satisfying $[L_{\sigma}(\lambda),H_{\mathrm{S}}(\lambda)]=0$, $L^\dag_{\sigma}(\lambda)=L_{\sigma}(\lambda)$, are related to pure dephasing with no energy relation. After further simplification, we have
\begin{equation}
\dot\rho_t=-\frac{i}{\hbar}[H(\lambda_t),\rho_t]+\sum_j\mathcal{D}[L_j(\lambda_t)]\rho_t,
\label{SALME}
\end{equation}
where $H(\lambda)=H_{\mathrm{S}}(\lambda)+H_{\mathrm{LS}}(\lambda)$, and $L_j(\lambda)$ ($j=ab$ or $j=\sigma$) satisfies $[L_j(\lambda),H_{\mathrm{S}}(\lambda)]=\Delta_j(\lambda)L_j(\lambda)$ and $L^\dag_{j'}(\lambda)=L_j(\lambda)e^{-\beta\Delta_j(\lambda)/2}$, with $\Delta_j$ the energy change of the $j$-th quantum jump. Since $H_{\rm LS}(\lambda)$ is usually negligible compared with $H_{\rm S}(\lambda)$, we simply neglect it and treat $H(\lambda)$ identically as $H_{\rm S}(\lambda)$ \footnote{Though such approximation is made in many textbooks and research papers, as is highlighted in Ref.~\cite{Lidar2012}, the Lamb shift may considerably modify the long time dynamics of the system, because it is typically of the same order of magnitude of the state transition rates. But to avoid a subtle problem that whether the Lamb shift Hamiltonian should be identified as inclusive or exclusive, we just ignore it for the time being.}.  As mentioned in the beginning, the mixed LME (\ref{EOM}) in the main text is obtained from the adiabatic LME (\ref{SALME}) if we simply add the perturbation $h_t$ into the unitary part.

\section{Remarks on quantum trajectory thermodynamics}
\label{RQTT}

In the weak-coupling regime, we can always interpret heat (work) as the energy exchange between the system and the heat bath (the total energy increment) \cite{Talkner2011}.  However, even in this regime, addressing work and heat is highly nontrivial for quantum systems and at the trajectory level, because a quantum system can generally be a superposition of energy eigenstates, and we cannot have an objective concept of ``trajectory" \cite{Wiseman2012}.

Fortunately, for isolated quantum systems, work coincides with the energy change, and a consensus has been achieved that the two-time energy measurement (TTEM) \cite{Kurchan2000,Tasaki2000,Talkner2007} gives the most reasonable definition of quantum work. Here a trajectory can be specified by the two outcomes $E^{\lambda_0}_a$ and $E^{\lambda_\tau}_b$ of the TTEM and the work is simply their subtraction $W=E^{\lambda_\tau}_b-E^{\lambda_0}_a$. The TTEM definition implies the Jarzynski equality (and hence the second law) and is experimentally relevant \cite{Quan2015,Paternostro2014}. Theoretically, the consistency between the TTEM definition of work and the classical counterpart has been proved for one-dimensional systems \cite{Jarzynski2015}.

Combining the idea of TTEM with the Hamiltonian formalism of classical nonequilibrium thermodynamics \cite{Jarzynski2011}, the joint TTEM approach was proposed to define work and heat for open quantum systems \cite{Esposito2009,Talkner2011}, where a trajectory is specified by the initial and final eigenenergies of the system $E^{\lambda_0}_a, E^{\lambda_\tau}_b$, and those of the heat bath $E^{\rm B}_i,E^{\rm B}_j$, with the work and heat being respectively $W=E^{\rm B}_j+E^{\lambda_\tau}_b-E^{\rm B}_i-E^{\lambda_0}_a$ and $Q=E^{\rm B}_j-E^{\rm B}_i$. To obtain stochastic thermodynamics from the deterministic Hamiltonian formalism, the detailed information of the heat bath should be traced out,  as is done in the classical case \cite{Sekimoto2010}. Using the characteristic function approach \cite{Talkner2007}, one can encode the statistics of work \cite{Silaev2014} and heat \cite{Esposito2009} into a generalized quantum master equation after the standard Born-Markov approximation and the rotating-wave approximation. Under such coarse-graining, the statistics of work and that of heat turn out to be consistent with the formalism in the main text \cite{Maes2008,Liu2016}.

Therefore, the QJT-based definition naturally emerges from the two facts that (i) heat (work) is the energy change of the heat bath (the system and the heat bath) and that (ii) the energy change is quantified by the TTEM. While deriving the QJT-based definition from the TTEM approach is rather technical, the work and heat along a QJT per se can be explained intuitively. According to the interpretation (continuous monitoring) of a QJT, if the $j$-th QJ is detected at time $t$, an energy quanta equal to $\Delta_j(\lambda_t)$ is transferred from the system to the heat bath; thus the accumulated heat should increase by $\Delta_j(\lambda_t)$. For example, in a photodetection experiment where a two-level atom with a constant energy gap $\Delta$ interacts with the photon field in an optical cavity,  the heat along a QJT is the \emph{net} number of the photons \emph{emitted} by the atom multiplied by $\Delta$ in a single experimental realization \cite{Pekola2013}. Once the heat along a QJT is obtained, the work can be determined by the first law of thermodynamics, as mentioned in the main text.

The continuous monitoring interpretation of a QJT can be heuristically shown as follows. A QJ operator $L_{ab}(\lambda_t)$ with a nonzero energy effect, which corresponds to a state transition, is actually the sum of all the operators $\langle e_j|e^{-\frac{i}{\hbar}H_{\rm tot}(t)\delta t}|e_i\rangle e^{-\beta E^{\rm B}_i}/(\delta t Z_{\rm B})$ in Ref.~\cite{Crooks2008} with the same energy difference $E^{\rm B}_j-E^{\rm B}_i=E^{\lambda_t}_a-E^{\lambda_t}_b\neq0$ ($H_{\rm B}|e_j\rangle=E^{\rm B}_j|e_j\rangle$) under the first-order perturbation theory, namely Fermi's golden rule. This connection is clear in the Lehmann representation of $w_{ba}(\lambda)$ in Eq.~(\ref{Lab}):
\begin{equation}
\begin{split}
w_{ba}(\lambda)&=\frac{2\pi g^2}{\hbar}\sum_{i,j}|\langle a^\lambda,e_j|\sum_\alpha A_\alpha\otimes B_\alpha|b^\lambda,e_i\rangle|^2\\
&\times\frac{e^{-\beta E^{\rm B}_i}}{Z_{\rm B}}\delta(E^{\rm B}_j+E^\lambda_a-E^{\rm B}_i-E^\lambda_b).
\end{split}
\end{equation}
More accurately, to achieve the summation $\sum_{i,j}$ in a single experimental realization, we should apply the so-called generalized quantum measurement \cite{Paz2014} 
instead of the usual two-time projective measurement approach \cite{Talkner2011}. On the other hand, the detection of a dephasing QJ $L_\sigma(\lambda)$ seems hard to be implemented even in principle, since there is no energy exchange between the system and the heat bath, which makes no difference from each other or from the nonunitary evolution. For mathematical reasons \footnote{Actually a LME $\dot\rho_t=-\frac{i}{\hbar}[H_t,\rho_t]+\sum_j\mathcal{D}[L^j_t]\rho_t$ can be rewritten into $\dot\rho_t=-\frac{i}{\hbar}[H_t,\rho_t]+\sum_j\mathcal{D}[L^{\star j}_t]\rho_t$ for an arbitrary unitary transformation between the jump operators $L^{\star j}_t=\sum_k u_{jk}(t)L^k_t$. However, an instantaneously detailed-balanced LME always has a set of privileged jump operators $L^j_t$ satisfying $[L^j_t,\ln\pi_t]=\Phi^j_t L^j_t$, with $\pi_t$ being the instantaneous equilibrium state and $\Phi^j_t$'s being c-numbers (see Refs.~\cite{Horowitz2013,Horowitz2015} and the references therein). Therefore, from a mathematical point of view, it is natural to teat $L_\sigma(\lambda_t)$ and $L_{ab}(\lambda_t)$ identically, since they constitute such a privileged set.}, and to be consistent with the generalized master equation approach \cite{Esposito2009,Silaev2014,Liu2016} and the quantum Feynman-Kac formula method-based \cite{Liu2012,Chetrite2012,Liu2014a} definition of work and heat distributions, we treat $L_\sigma(\lambda)$ in the same manner as $L_{ab}(\lambda)$ in the general formalism. However this is controversial. For example, Ref.~\cite{Liu2014b} treated $L_{\sigma}(\lambda_t)$ as a QJ, while in Ref.~\cite{Pekola2015a} it is unravelled as quantum diffusion. To avoid the ambiguity in an experiment-relevant model, we choose an example in which all the diagonal (in the energy representation) matrix elements of $A_\alpha$ and $B_\alpha$ vanish so that there is no dephasing QJ, just like Refs.~\cite{Horowitz2012,Pekola2013}.

Finally, it is worth comparing our formalism with a different quantum trajectory-based framework for stochastic thermodynamics established quite recently \cite{Elouard2015,Lutz2016}. In that framework, the change of the energy \emph{expectation} due to the deterministic (stochastic) part of the change of the state $\tilde\rho_t$ in a single realization, which is \emph{not} necessarily pure due to the imperfect continuous monitoring, is identified as work (heat), namely
\begin{equation}
\begin{split}
W[\tilde\rho_t]&=\int^\tau_0 dt\dot\lambda_t \mathrm{Tr}[\partial_\lambda H(\lambda_t)\tilde\rho_t],\\
Q[\tilde\rho_t]&=\mathrm{Tr}[H(\lambda_0)\tilde\rho_t]-\mathrm{Tr}[H(\lambda_\tau)\tilde\rho_t]\\
&+\int^\tau_0 dt\dot\lambda_t \mathrm{Tr}[\partial_\lambda H(\lambda_t)\tilde\rho_t].
\end{split}
\end{equation}
This formalism also allows intuitive physical interpretation, and is clearly consistent with quantum thermodynamics at the ensemble level. Moreover, this formalism is applicable to any kind of unravelling, such as the quantum diffusion mentioned before \cite{Lutz2016}, while our formalism no longer works for the systems where the rotating-wave approximation is invalid (e.g., quantum Brownian motion \cite{Breuer2002,Wiseman2010}). On the other hand, this formalism cannot reproduce the widely accepted TTEM definition in the adiabatic limit, and, as a result, does not imply the fluctuation theorems or the second law \cite{Lutz2016}. In contrast, several fluctuation theorems have been derived within our framework \cite{Horowitz2012,Pekola2013,Horowitz2013}. Therefore, in the context of nonequilibrium fluctuation theory, our framework should be the better choice.

\section{Consistency at the ensemble level and in the classical or adiabatic limit}
\label{Consistency}
\subsection{Ensemble level}
We first consider the case without feedback control. Consider a small time interval $[t,t+dt]$ during which the probability that the $j$-th quantum jump occurs at the ensemble level is
\begin{equation} 
\begin{split}
&\mathrm{E}[dN^j_t]=\mathrm{E}[\langle\psi_t|L^\dag_j(\lambda_t)L_j(\lambda_t)|\psi_t\rangle dt]\\
&=\mathrm{Tr}[L^\dag_j(\lambda_t)L_j(\lambda_t)\mathrm{E}[|\psi_t\rangle\langle\psi_t|]dt]=\mathrm{Tr}[L^\dagger_j(\lambda_t)L_j(\lambda_t)\rho_t]dt,
\end{split}
\end{equation}
which is accompanied by a heat generation by the amount of $\Delta_j$. By multiplying the heat generation $\Delta_j$ due to this quantum jump and then summing up all the dissipation and dephasing channel indexes $j$, we obtain the averaged heat accumulated during such a small time interval as
\begin{widetext}
\begin{equation}
\begin{split}
&d\langle Q\rangle_t =\sum_j \mathrm{Tr}[L_j(\lambda_t)\rho_t L^\dagger_j(\lambda_t)]\Delta_j(\lambda_t)dt=\sum_j \mathrm{Tr}[L^\dagger_j(\lambda_t)[H(\lambda_t)+\Delta_j(\lambda_t)]L_j(\lambda_t)\rho_t-H(\lambda_t)L_j(\lambda_t)\rho_t L^\dagger_j(\lambda_t)]dt\\
&=\sum_j\mathrm{Tr}[H(\lambda_t)[L^\dagger_j(\lambda_t)L_j(\lambda_t)\rho_t-L_j(\lambda_t)\rho_tL^\dagger_j(\lambda_t)]]dt
=-\mathrm{Tr}[H(\lambda_t)\sum_j\mathcal{D}[L_j(\lambda_t)]\rho_t]dt\\
&=-\mathrm{Tr}\left[H(\lambda_t)\left(\frac{i}{\hbar}[H(\lambda_t)+h_t,\rho_t]+\dot\rho_t\right)\right]dt=-\mathrm{Tr}[H(\lambda_t)\dot\rho_t]dt-\frac{i}{\hbar}\mathrm{Tr}[H(\lambda_t)[h_t,\rho_t]]dt.
\end{split}
\end{equation}
\end{widetext}
Here we have used $[L_j(\lambda),H(\lambda)]=\Delta_j(\lambda) L_j(\lambda)$ (so that $[L^\dag_j(\lambda)L_j(\lambda),H(\lambda)]=0$), $\mathrm{Tr}[AB]=\mathrm{Tr}[BA]$ and the LME (\ref{EOM}). By using the first law of thermodynamics at the ensemble level, we finally obtain
\begin{equation}
\begin{split}
&d\langle W\rangle_t=d\langle H(\lambda_t)\rangle_t+d\langle Q\rangle_t\\
&=\mathrm{Tr}[\partial_\lambda H(\lambda_t)\rho_t]\dot\lambda_tdt+\frac{i}{\hbar}\mathrm{Tr}[[h_t,H(\lambda_t)]\rho_t]dt.
\end{split}
\end{equation}
Therefore, the total averaged heat and work during the process are given by Eq.~(\ref{QWQ}).

In the present of feedback control,
to carefully identify the energy effect of the measurement, we had better start from the original definition of the heat at the trajectory level, instead of inadvertently applying Eq.~(\ref{QWQ}). We can easily find the problem that the averaged heat production $\delta \langle Q\rangle_{\rm m}=-\sum_j \mathrm{Tr}[L_j(\lambda_{t_{\rm m}})\rho_{t_{\rm m}} L^\dagger_j(\lambda_{t_{\rm m}})]\Delta_j(\lambda_{t_{\rm m}})\delta t_{\rm m}$ during $[t_{\rm m}-\delta t_{\rm m}/2,t_{\rm m}+\delta t_{\rm m}/2]$ is ill-defined, because $\rho_{t_{\rm m}}$ is indeterminable.
This problem arises from the idealized assumption that the measurement takes place instantaneously, and thus can be solved by quantifying the Hamiltonians of the measurement device and its interaction with the system for a finite $\delta t_{\rm m}$. Nevertheless, $\delta \langle Q\rangle_{\rm m}$ should be of the order of $O(\delta t_{\rm m}/\tau)$ compared with the total averaged heat, since $\delta \langle Q\rangle_{\rm m}$ is roughly proportional to the density operator (always bounded) rather than its time derivative. Therefore, we can safely neglect $\delta \langle Q\rangle_{\rm m}$ in the $\delta t_{\rm m}\to 0$ limit and evaluate the total heat as
\begin{widetext}
\begin{equation}
\langle Q\rangle=-\sum_\alpha p_\alpha\int^{\tau}_{t^+_{\rm m}}dt\left(\mathrm{Tr}[H(\lambda^\alpha_t)\dot\rho^\alpha_t]+\frac{1}{i\hbar}\mathrm{Tr}[[h^\alpha_t,H(\lambda^\alpha_t)]\rho^\alpha_t]\right)-\int^{t^-_{\rm m}}_0 dt\left(\mathrm{Tr}[H(\lambda_t)\dot\rho_t]+\frac{1}{i\hbar}\mathrm{Tr}[[h_t,H(\lambda_t)]\rho_t]\right),
\end{equation}
\end{widetext}
where $\rho^\alpha_t$ ($t>t_{\rm m}$) is the solution to
\begin{equation}
\dot\rho_t=-\frac{i}{\hbar}[H(\lambda^\alpha_t)+h^\alpha_t,\rho_t]+\sum_j\mathcal{D}[L_j(\lambda^\alpha_t)]\rho_t
\end{equation}
for the initial condition $\rho^\alpha_{t^+_{\rm m}}=M_\alpha \rho_{t^-_{\rm m}}M^\dag_\alpha/p_\alpha$, $p_\alpha=\mathrm{Tr}[M^\dag_\alpha M_\alpha\rho_{t^-_{\rm m}}]$ corresponding to the selective post-measurement state. Accordingly, the total averaged work $\langle W\rangle=\langle H(\lambda^\alpha_\tau)\rangle_{\tau,\alpha}-\langle H(\lambda_0)\rangle_0+\langle Q\rangle$ reads
\begin{widetext}
\begin{equation}
\begin{split}
&\langle W\rangle=\sum_\alpha p_\alpha\int^{\tau}_{t^+_{\rm m}}dt\left(\dot\lambda^\alpha_t\mathrm{Tr}[\partial_\lambda H(\lambda^\alpha_t)\rho^\alpha_t]-\frac{1}{i\hbar}\mathrm{Tr}[[h^\alpha_t,H(\lambda^\alpha_t)]\rho^\alpha_t]\right)\\
&+\int^{t^-_{\rm m}}_0 dt\left(\dot\lambda_t\mathrm{Tr}[\partial_\lambda H(\lambda_t)\dot\rho_t]-\frac{1}{i\hbar}\mathrm{Tr}[[h_t,H(\lambda_t)]\rho_t]\right)+\mathrm{Tr}[H(\lambda_{t_{\rm m}})(\rho_{t^+_{\rm m}}-\rho_{t^-_{\rm m}})],
\end{split}
\label{workexp}
\end{equation}
\end{widetext}
where $\rho_{t^+_{\rm m}}=\sum_\alpha M_\alpha\rho_{t^-_{\rm m}} M^\dag_\alpha=\sum_\alpha p_\alpha\rho^\alpha_{t^+_{\rm m}}$ is the nonselective postmeasurement state. One can see that the last term in Eq.~(\ref{workexp}) corresponds to the energy change of the system induced by the measurement backaction. Thus we have confirmed that the quantum trajectory thermodynamics does reduce to the conventional quantum thermodynamics at the ensemble level irrespective of the presence of feedback control.

\subsection{Classical limit}
A LME with a \emph{time-independent} generator can be decoupled to a classical Markovian (Pauli) master equation 
of the diagonal elements of the density matrix, and a set of independent dephasing equations of the off-diagonal elements \cite{Breuer2002}. While in the time-dependent case, the noncommutativity of $H(\lambda)$ with different work parameters $\lambda$ and that with $h_t$ lead to quantum tunneling between different instantaneous eigenstates, thereby coupling the time evolution of the diagonal and the off-diagonal density matrix elements. This makes the dynamics, and consequently thermodynamics, very complicated. However, if the noncommutativity is negligible, which we call the classical limit and is achievable for an extremely slow driving or a special kind of $H(\lambda)$ whose eigenstates are independent of $\lambda$, the system becomes classical and the dynamics is described by the time-dependent Pauli master equation 
\begin{equation}
\dot p_b(t)=\sum_a [w_{ab}(\lambda_t)p_a(t)-w_{ba}(\lambda_t)p_b(t)],
\label{PauliE}
\end{equation}
where $p_a(t)\equiv\langle a^{\lambda_t}|\rho_t|a^{\lambda_t}\rangle$. Equation (\ref{PauliE}) should be sufficient for the description of the dynamics as long as the initial state only has nonzero diagonal elements (
e.g., the equilibrium state). We will show that the quantum trajectory thermodynamics 
recovers the well-established classical stochastic thermodynamics in the Pauli master equation formalism \cite{Seifert2008}. 

For simplicity, we arrive at the classical limit by assuming $[H(\lambda),H(\lambda')]=[H(\lambda),h_t]=0$, so that $H(\lambda)=\sum_n E_n(\lambda)|n\rangle\langle n|$ with $|n\rangle$ being $\lambda$-independent. The 
system undergoes (nonunitary) quantum adiabatic evolution, no matter how sensitively $E_n(\lambda)$ depends on $\lambda$ during any two QJs. In this case, a QJT $\psi_t$ with a nonzero probability must be like
\begin{widetext}
\begin{equation}
\begin{split}
\psi_t:\;m_0 \xrightarrow{d^{m_0}_{j_{01}}(\lambda_{t_{01}})} m_0 \xrightarrow{d^{m_0}_{j_{02}}(\lambda_{t_{02}})}m_0...m_0 \xrightarrow{d^{m_0}_{j_{0r_0}}(\lambda_{t_{0r_0}})} m_0 \xrightarrow{w_{m_0m_1}(\lambda_{t_1})} m_1 \xrightarrow{d^{m_1}_{j_{11}}(\lambda_{t_{11}})} m_1...m_1 \xrightarrow{d^{m_1}_{j_{1r_1}}(\lambda_{t_{1r_1}})} m_1 \\
\xrightarrow{w_{m_1m_2}(\lambda_{t_2})} m_2 ... m_{M-1}\xrightarrow{w_{m_{M-1}m_M}(\lambda_{t_M})} m_M \xrightarrow{d^{m_M}_{j_{M1}}(\lambda_{t_{M1}})} m_M...m_M \xrightarrow{d^{m_M}_{j_{Mr_M}}(\lambda_{t_{Mr_M}})} m_M,
\end{split}
\label{adiabaticQT}
\end{equation}
\end{widetext}
where only QJs are presented, with $w_{m_pm_{p+1}}(\lambda_{t_p})$ ($d^{m_p}_{j_{pq}}(\lambda_{t_{pq}})$) denoting a state transition (dephasing) QJ with nonzero (zero) heat production. Owing to the quantum adiabatic evolution that maintains the quantum number, 
such a QJT is very similar to a classical one except for the dephasing QJs. The heat (work) along this QJT are completely determined by the state transition QJs and the initial and the final state energies:
\begin{equation}
\begin{split}
&Q[\psi_t]=\sum^{M-1}_{k=0} (E^{\lambda_{t_{k+1}}}_{m_k}-E^{\lambda_{t_{k+1}}}_{m_{k+1}}),\\
&W[\psi_t]=\sum^M_{k=0} (E^{\lambda_{t_{k+1}}}_{m_k}-E^{\lambda_{t_k}}_{m_k}),
\end{split}
\label{CWQ}
\end{equation}
where $t_0\equiv0$ and $t_{M+1}\equiv\tau$.

In fact, we can figure out the exact probability of a classical trajectory if we sum over all the QJTs with the same classical reduction. To do this, we first define the classical reduction $m_t$ of a QJT $\psi_t$ (\ref{adiabaticQT}) 
as follows:
\begin{equation}
\begin{split}
m_t=\mathfrak{R}[\psi_t]:\;&m_0 \xrightarrow{w_{m_0m_1}(\lambda_{t_1})} m_1 \xrightarrow{w_{m_1m_2}(\lambda_{t_2})} m_2\\
...&m_{M-1}\xrightarrow{w_{m_{M-1}m_M}(\lambda_{t_M})} m_M,
\end{split}
\label{CT}
\end{equation}
where only the state transition QJs are retained. Such a definition is reasonable because the classical work and heat along the reduced classical trajectory (\ref{CT}) are defined by Eq.~(\ref{CWQ}) \cite{Crooks1998}. For convenience but without the loss of generality, we denote $L_1(\lambda),L_2(\lambda),...,L_{J_1}(\lambda)$ as all the dephasing jump operators, each of which takes the form $L_j(\lambda)=\sum_n d^n_j(\lambda)|n\rangle\langle n|$ (since $[L_j(\lambda),H(\lambda)]=0$). The remaining state transition jump operators must take the form of $L_{ab}(\lambda)=\sqrt{w_{ba}(\lambda)}|a\rangle \langle b|$. Now we write down the conditional probability of the QJT $\psi_t$ as 
\begin{widetext}
\begin{equation}
\mathcal{P}[\psi_t|\psi_0] = \prod^{M-1}_{p=0}w_{m_pm_{p+1}}(\lambda_{t_{p+1}})dt_{p+1}\times\prod^M_{p=0} e^{-\int^{t_{p+1}}_{t_p}dt [w_{m_p}(\lambda_t) + D_{m_p}(\lambda_t)]}\prod^{r_p}_{q=1}|d^{m_p}_{j_{pq}}(\lambda_{t_{pq}})|^2 dt_{pq},
\end{equation}
\end{widetext}
where $t_0\equiv0$, $t_{M+1}\equiv \tau$, $j_{pq}\in\{1,2,...,J_1\}$, $w_n(\lambda)\equiv\sum_{m\ne n}w_{nm}(\lambda)$ and $D_n(\lambda)\equiv\sum^{J_1}_{j=1}|d^n_j(\lambda)|^2$. Then we sum up the conditional probabilities of all the $\psi_t$ corresponding to the same $m_t$, leading to
\begin{widetext}
\begin{equation}
\begin{split}
\mathcal{P}[m_t|m_0] = \int_{\{\psi_t:\mathfrak{R} [\psi_t]=m_t\}} D[\psi_t] \mathcal{P}[\psi_t|\psi_0] = \prod^{M-1}_{p=0}w_{m_pm_{p+1}}(\lambda_{t_{p+1}})dt_{p+1} \cdot \prod^M_{p=0} e^{-\int^{t_{p+1}}_{t_p}dt [w_{m_p}(\lambda_t) + D_{m_p}(\lambda_t)]} \\ \times\sum^{+\infty}_{r_p=0}\int^{t_{p+1}}_{t_p}dt_{pr_p} ... \int^{t_{p3}}_{t_p} dt_{p2} \int^{t_{p2}}_{t_p} dt_{p1}
\prod^{r_p}_{q=1} \left(\sum^{J_1}_{j_{pq}=1}|d^{m_p}_{j_{pq}}(\lambda_{t_{pq}})|^2\right).
\end{split}
\end{equation}
\end{widetext}
By using the identity $e^{\int^{t''}_{t'}dtf(t)}=\sum^{+\infty}_{r=0}\int^{t''}_{t'}dt_r...\int^{t_3}_{t'}dt_2\int^{t_2}_{t'}dt_1\prod^r_{q=1} f(t_q)$, we finally obtain
\begin{equation}
\begin{split}
&\mathcal{P}[m_t|m_0] = e^{-\int^\tau_{t_M}dt w_{m_M}(\lambda_t)}\\
&\times\prod^{M-1}_{p=0}w_{m_pm_{p+1}}(\lambda_{t_{p+1}})dt_{p+1}e^{-\int^{t_{p+1}}_{t_p}dt w_{m_p}(\lambda_t)},
\end{split}
\end{equation}
which turns out to be consistent with the conditional probability of a classical trajectory \cite{Esposito2010}. The generalization to the case with feedback control is straightforward, since there is no measurement backaction in the classical case.

It is worth mentioning that if $h_t$ generates a sudden permutation operation between different classical states, the exclusive driving can stay classical but perform nonzero work. Such an operation routinely occurs in a classical computer as in the reversible classical logic gate operation of classical bits.

\subsection{Adiabatic limit}
To reach the adiabatic limit, we only have to set $g=0$, so the system is dissipation-free and undergoes unitary evolution governed by the Liouville-von Neumann equation $\dot\rho_t=-\frac{i}{\hbar}[H(\lambda_t)+h_t,\rho_t]$. The QJT in this case is very simple: it only consists of the initial and final PM outcomes, while no QJ occurs, leading to $Q[\psi_t]=0$ and $W[\psi_t]=E^{\lambda_\tau}_b-E^{\lambda_0}_a$ which is the widely accepted two-time PM definition of quantum work in isolated quantum systems \cite{Tasaki2000}. When there is feedback, the energy change contributed by the measurement backaction is identified as work because of $\langle W\rangle=\langle\Delta E\rangle$ and $Q[\psi_t]$ always vanishes.

\section{Derivations and discussions of the generalized Jarzynski equalities}
\label{DDGJE}
\subsection{Derivation of Eq.~(\ref{TrajDB})}
\label{DTDB}
A QJT in a feedback control process can be completely characterized by a discrete set of outcomes $a$ and $b$ of the initial and the final PMs, the outcome $\alpha$ of the measurement $\rm M_A$, the total number of QJs $K$, and the time $t_k$ and the type $j_k$ of the $k$-th QJ. Given these parameters and a set of time resolutions $dt_k$, the probability of this forward QJT $\psi_t$ follows the stochastic Schr\"odinger equation (\ref{SSE}) and is given by
\begin{widetext}
\begin{equation}
\begin{split}
&\mathcal{P}[\psi_t,\alpha]=|\langle b^{\lambda^\alpha_\tau}| \left[\prod^K_{k=K_{\rm m}+1}U^\alpha_{\mathrm{eff}}(t_{k+1},t_k)L_{j_k}(\lambda^\alpha_{t_k})\right]U^\alpha_{\mathrm{eff}}(t_{K_{\rm m}+1},t_{\rm m})\\ 
&\times M_\alpha U_{\mathrm{eff}}(t_{\rm m},t_{K_{\rm m}}) \left[\prod^{K_{\rm m}}_{k=1}L_{j_k}(\lambda_{t_k})U_{\mathrm{eff}}(t_k,t_{k-1})\right] |a^{\lambda_0}\rangle|^2 p^{\rm eq}_a(\lambda_0)\prod^K_{k=1} dt_k,
\end{split}
\end{equation}
\end{widetext}
where 
$p^{\rm eq}_a(\lambda)\equiv e^{-\beta E^\lambda_a}/Z^\lambda$ is the probability that the system at the $a$-th eigenstate for the canonical ensemble with work parameter $\lambda$, and $U_{\mathrm{eff}}(t,t')$ ($U^\alpha_{\mathrm{eff}}(t,t')$) is the \emph{nonunitary} effective time-evolution operator generated by $H_{\mathrm{eff}}(t)$ ($H^\alpha_{\mathrm{eff}}(t)$
). Based on the definition of the corresponding time-reversed QJT $\bar\psi_t$ (see Fig.~1 in the main text), we can write down its probability as
\begin{widetext}
\begin{equation}
\begin{split}
&\mathcal{\bar P}[\bar\psi_t,\alpha]=|\langle a^{\bar\lambda_\tau}|\Theta^\dag \left[\prod^K_{k=\bar K_m+1}\bar U_{\mathrm{eff}}(\bar t_{k+1},\bar t_k)\bar L_{\bar j_k}(\bar\lambda_{\bar t_k})\right]\bar U_{\mathrm{eff}}(\bar t_{\bar K_{\rm m}+1},\bar t_{\rm m})\\ 
&\times \tilde M_\alpha \bar U^\alpha_{\mathrm{eff}}(\bar t_{\rm m},\bar t_{\bar K_{\rm m}}) \left[\prod^{\bar K_{\rm m}}_{k=1}\bar L_{\bar j_k}(\bar\lambda^\alpha_{\bar t_k})\bar U^\alpha_{\mathrm{eff}}(\bar t_k,\bar t_{k-1})\right] \Theta|b^{\bar\lambda^\alpha_0}\rangle|^2p^{\rm eq}_b(\bar\lambda^\alpha_0) J_K \prod^K_{k=1} d\bar t_k,
\end{split}
\end{equation}
\end{widetext}
where $J_K=(-)^K$ is the Jacobian $\partial(t_1,t_2,...,t_K)/\partial(\bar t_1,\bar t_2,...,\bar t_K)$, $t_0\equiv0$, $t_{K+1}\equiv\tau$, $\bar\lambda_t\equiv\lambda_{\tau-t}$ ($\bar\lambda^\alpha_t\equiv\lambda^\alpha_{\tau-t}$), $\bar t_{k}\equiv\tau-t_{K+1-k}$, $\bar t_{\rm m}\equiv\tau-t_{\rm m}$, $\bar K_{\rm m}\equiv K-K_{\rm m}$, $\bar j_k\equiv j'_{K+1-k}$ (we recall that $j'$ is uniquely determined by $\Delta_{j'}(\lambda)=-\Delta_j(\lambda)$ if $\Delta_j(\lambda)\neq0$ and $j'=j$ otherwise), $\tilde M_\alpha=\Theta M^\dag_\alpha\Theta^\dag$ and $\bar U_{\mathrm{eff}}(t,t')$ is generated by $\bar H_{\mathrm{eff}}(t)=\bar H(\bar\lambda_t)+\bar h_t-\sum_ji\hbar \bar L^\dag_j(\bar\lambda_t)\bar L_j(\bar\lambda_t)/2$, with $\bar O\equiv\Theta O\Theta^\dag$ ($\bar O_t\equiv\Theta O_{\tau-t}\Theta^\dag$) if $O$ is not explicitly time-dependent (if $O$ has a time argument).  One can show $\bar H_{\rm eff}(t)=\Theta H^\dag_{\rm eff}(\tau-t)\Theta^\dag$, which leads to $\bar U_{\mathrm{eff}}(t,t')=\Theta U^\dag_{\mathrm{eff}}(\tau-t',\tau-t)\Theta^\dag$ ($\bar U^\alpha_{\mathrm{eff}}(t,t')=\Theta U^{\alpha\dag}_{\mathrm{eff}}(\tau-t',\tau-t)\Theta^\dag$) \cite{Horowitz2013}. By substituting all these expressions into $\mathcal{\bar P}[\bar\psi_t,\alpha]$, we obtain
\begin{widetext}
\begin{equation}
\begin{split}
&\mathcal{\bar P}[\bar\psi_t,\alpha]=|\langle a^{\lambda_0}|\left[\prod^K_{k=K- K_{\rm m}+1} U^\dag_{\mathrm{eff}}(\tau-\bar t_{k},\tau-\bar t_{k+1}) L_{j'_{K-k+1}}(\lambda_{\tau-\bar t_k})\right] U^\dag_{\mathrm{eff}}(\tau-\bar t_{\rm m},\tau-\bar t_{K-K_{\rm m}+1})\\ 
&\times M^\dag_\alpha U^{\alpha\dag}_{\mathrm{eff}}(\tau-\bar t_{K-K_{\rm m}},\tau-\bar t_{\rm m}) \left[\prod^{K-K_{\rm m}}_{k=1} L_{j'_{K-k+1}}(\lambda^\alpha_{\tau-\bar t_k}) U^{\alpha\dag}_{\mathrm{eff}}(\tau-\bar t_{k-1},\tau-\bar t_k)\right]|b^{\lambda^\alpha_\tau}\rangle|^2 p^{\rm eq}_b(\lambda^\alpha_\tau) J_K \prod^K_{k=1} d\bar t_k \\
&=|\langle a^{\lambda^\alpha_\tau}|\left[\prod^K_{k=K- K_{\rm m}+1} U^\dag_{\mathrm{eff}}(t_{K-k+1},t_{K-k}) L^\dag_{j_{K-k+1}}(\lambda_{t_{K-k+1}})e^{-\frac{1}{2}\beta\Delta_{j_{K-k+1}}(\lambda_{t_{K-k+1}})}\right] U^\dag_{\mathrm{eff}}(t_{\rm m},t_{K_{\rm m}})M^\dag_\alpha \\ 
&\times U^{\alpha\dag}_{\mathrm{eff}}(t_{K_{\rm m}+1},t_{\rm m})\left[\prod^{K-K_{\rm m}}_{k=1} L^\dag_{j_{K-k+1}}(\lambda^\alpha_{t_{K-k+1}})e^{-\frac{1}{2}\beta\Delta_{j_{K-k+1}}(\lambda^\alpha_{t_{K-k+1}})} U^{\alpha\dag}_{\mathrm{eff}}(t_{K-k+2},t_{K-k+1})\right]|b^{\lambda_0}\rangle|^2p^{\rm eq}_b(\lambda^\alpha_\tau) \prod^K_{k=1} dt_k \\
&=|\langle a^{\lambda^\alpha_\tau}|\left[\prod^{K_{\rm m}}_{k=1} U^\dag_{\mathrm{eff}}(t_k,t_{k-1}) L^\dag_{j_k}(\lambda_{t_k})e^{-\frac{1}{2}\beta\Delta_{j_k}(\lambda_{t_k})}\right] U^\dag_{\mathrm{eff}}(t_{\rm m},t_{K_{\rm m}})M^\dag_\alpha U^{\alpha\dag}_{\mathrm{eff}}(t_{K_{\rm m}+1},t_{\rm m})\\ 
&\times \left[\prod^{K}_{k=K_{\rm m}+1} L^\dag_{j_k}(\lambda^\alpha_{t_k})e^{-\frac{1}{2}\beta\Delta_{j_k}(\lambda^\alpha_{t_k})} U^{\alpha\dag}_{\mathrm{eff}}(t_{k+1},t_k)\right]|b^{\lambda_0}\rangle|^2p^{\rm eq}_b(\lambda^\alpha_\tau) \prod^K_{k=1} dt_k \\
&=e^{-\beta[E^{\lambda^\alpha_\tau}_b-E^{\lambda_0}_a+\sum^K_{k=1}\Delta_{j_k}(\lambda^\alpha_{t_k})-\Delta F_\alpha]}|\langle b^{\lambda^\alpha_\tau}| \left[\prod^K_{k=K_{\rm m}+1}U^\alpha_{\mathrm{eff}}(t_{k+1},t_k)L_{j_k}(\lambda^\alpha_{t_k})\right]U^\alpha_{\mathrm{eff}}(t_{K_{\rm m}+1},t_{\rm m})M_\alpha \\ 
&\times U_{\mathrm{eff}}(t_{\rm m},t_{K_{\rm m}}) \left[\prod^{K_{\rm m}}_{k=1}L_{j_k}(\lambda_{t_k})U_{\mathrm{eff}}(t_k,t_{k-1})\right] |a^{\lambda_0}\rangle|^2p^{\rm eq}_a(\lambda_0)\prod^K_{k=1} dt_k=e^{-\beta(W[\psi_t,\alpha]-\Delta F_\alpha)}\mathcal{P}[\psi_t,\alpha],
\end{split}
\end{equation}
\end{widetext}
where $\lambda^\alpha_t\equiv\lambda_t$ for $t<t_{\rm m}$. Thus we have completed the proof of Eq.~(\ref{TrajDB}) in the main text
\begin{equation}
\mathcal{\bar P}[\bar\psi_t,\alpha]=e^{-\beta(W[\psi_t,\alpha]-\Delta F_\alpha)}\mathcal{P}[\psi_t,\alpha].
\label{DBTL}
\end{equation}

\subsection{Derivation of the efficacy of feedback control (\ref{EFC})}
\label{DEFC}
Using Eq.~(\ref{DBTL}), we have
\begin{widetext}
\begin{equation}
\langle e^{-\beta(W-\Delta F)}\rangle=\sum_\alpha\int D[\psi_t]\mathcal{P}[\psi_t,\alpha]e^{-\beta(W[\psi_t,\alpha]-\Delta F_\alpha)}=\sum_\alpha\int D[\bar \psi_t]\mathcal{\bar P}[\bar\psi_t,\alpha],
\label{GJE1p}
\end{equation}
\end{widetext}
where
\begin{widetext}
\begin{equation}
\begin{split}
\int D[\psi_t]\equiv\sum_{a,b}\sum^\infty_{K=0}\sum_{\{j_k:1\le k\le K\}}\prod^K_{k=1}\int^{t_{k+1}}_0 (\mathrm{with\;respect\;to\;} dt_k),\\
\int D[\bar\psi_t]\equiv\sum_{b,a}\sum^\infty_{K=0}\sum_{\{\bar j_k:1\le k\le K\}}\prod^K_{k=1}\int^{\bar t_{k+1}}_0 J^{-1}_K (\mathrm{with\;respect\;to\;} d\bar t_k).
\end{split}
\end{equation}
\end{widetext}
Then we calculate the path integral involved on the right-hand side of Eq.~(\ref{GJE1p}) for a given $\alpha$
\begin{widetext}
\begin{equation}
\begin{split}
&\int D[\bar \psi_t]\mathcal{\bar P}[\bar\psi_t,\alpha]=\sum_{b,a}\sum^\infty_{\bar K_{\rm m}=0}\sum^\infty_{K=\bar K_{\rm m}}\sum_{\{\bar j_k:1\le k\le K\}}\left[\prod^{K}_{k=\bar K_{\rm m}+1}\int^{\bar t_{k+1}}_{\bar t_{\rm m}} d\bar t_k\right]\int^{\bar t_{\rm m}}_0d\bar t_{\bar K_{\rm m}}\prod^{\bar K_{\rm m}-1}_{k=1}\int^{\bar t_{k+1}}_0 d\bar t_k\; p^{\rm eq}_b(\bar\lambda^\alpha_0)\\
&\times|\langle a^{\bar\lambda_\tau}|\Theta^\dag \left[\prod^K_{k=\bar K_m+1}\bar U_{\mathrm{eff}}(\bar t_{k+1},\bar t_k)\bar L_{\bar j_k}(\bar\lambda_{\bar t_k})\right]\bar U_{\mathrm{eff}}(\bar t_{\bar K_{\rm m}+1},\bar t_{\rm m})\tilde M_\alpha \bar U^\alpha_{\mathrm{eff}}(\bar t_{\rm m},\bar t_{\bar K_{\rm m}}) \left[\prod^{\bar K_{\rm m}}_{k=1}\bar L_{\bar j_k}(\bar\lambda^\alpha_{\bar t_k})\bar U^\alpha_{\mathrm{eff}}(\bar t_k,\bar t_{k-1})\right] \Theta|b^{\bar\lambda^\alpha_0}\rangle|^2 \\
&=\mathrm{Tr}[\mathcal{T}_+ e^{\int^{\tau}_{\bar t_{\rm m}} dt \mathcal{\bar L}_t} \tilde M_\alpha[\mathcal{T}_+ e^{\int^{\bar t_{\rm m}}_0 dt \mathcal{\bar L}^\alpha_t}\bar\rho^{\rm eq}(\lambda^\alpha_\tau)] \tilde M^\dag_\alpha]=\mathrm{Tr}[\tilde M^\dag_\alpha\tilde M_\alpha\rho^\alpha_{\bar t^-_{\rm m}}],
\end{split}
\label{TRPI}
\end{equation}
\end{widetext}
where $\bar\rho^\alpha_t$ is the solution to $\dot\rho_t=\mathcal{\bar L}^\alpha_t\rho_t$ starting from the equilibrium state $\bar\rho^{\rm eq}(\lambda^\alpha_\tau)=e^{-\beta\bar H(\lambda^\alpha_\tau)}/Z^{\lambda^\alpha_\tau}$, and several properties have been used, including the trace preserving property of $\mathcal{L}_t$ and the path integral representation of the time evolution generated by a general time-dependent Lindblad-form superoperator $\mathcal{L}_t=-\frac{i}{\hbar}[H_t,\cdot]+\sum_j\mathcal{D}[L^j_t]\cdot$ 
\begin{equation}
\begin{split}
\mathcal{T}_+ e^{\int^{t''}_{t'} dt \mathcal{L}_t}&=\sum^\infty_{L=0}\sum_{\{j_l:1\le l\le L\}}\prod^L_{l=1}\int^{t_{l+1}}_{t'}dt_l\mathcal{U}_{\rm eff}(t'',t_L)\\
&\times\left[\prod^{L}_{l=1}\mathcal{J}_{j_l}(t_l)\mathcal{U}_{\rm eff}(t_l,t_{l-1})\right],
\end{split}
\end{equation}
where $t_0\equiv t'$ and $t_{L+1}\equiv t''$ for each summation term with definite $L$, and $\mathcal{J}_j(t)\rho\equiv L^j_t\rho L^{j\dag}_t$ and $\mathcal{U}_{\rm eff}(t,t')\rho=U_{\rm eff}(t,t')\rho U^\dag_{\rm eff}(t,t')$ are the jump \emph{superoperator} and the effective time-evolution superoperator, respectively. After substituting Eq.~(\ref{TRPI}) into Eq.~(\ref{GJE1p}), we finally come up with the first generalized Jarzynski equality:
\begin{equation}
\langle e^{-\beta(W-\Delta F)}\rangle=\sum_\alpha\mathrm{Tr}[\tilde M^\dag_\alpha\tilde M_\alpha\bar\rho^\alpha_{\bar t^-_{\rm m}}]=\eta_{\rm QJT}.
\end{equation}

The existence of a measurement $\mathrm{M}_{\mathrm{B}_\alpha}$ that contains $\tilde M_\alpha$ can be understood in the following manner. Based on either the picture of the system-measurement device interaction or a rigorous mathematical conclusion \cite{Zyczkowski2006}, we can express $M_\alpha$ as $M_\alpha=\langle \alpha_{\rm M}|U_{\rm SM}|\psi_{\rm M}\rangle$, and therefore $\tilde M_\alpha=\Theta\langle \psi_{\rm M}|U^\dag_{\rm SM}|\alpha_{\rm M}\rangle\Theta^\dag$. Starting from any given $|\psi_{\rm M}\rangle$, we can always find out another $D-1$ state vectors $|\phi^j_{\rm M}\rangle$, which can be made to satisfy $\langle \phi^j_{\rm M}|\phi^k_{\rm M}\rangle=\delta_{jk}$ and $\langle \phi^j_{\rm M}|\psi_{\rm M}\rangle=0$ ($j,k=1,2,...,D-1$) through the Schmidt orthogonalization process. Therefore, $\tilde M_\alpha$ and $\Theta\langle \phi^j_{\rm M}|U^\dag_{\rm SM}|\alpha_{\rm M}\rangle\Theta^\dag$ constitute a measurement $\mathrm{M}_{\mathrm{B}_\alpha}$. Here $D$ gives both the Hilbert-space dimension of the measurement device and the number of the measurement outcomes.

The consistency of $\eta_{\rm QJT}$ and the classical counterpart $\eta_{\rm C}$ \cite{Sagawa2010} in the classical limit can be understood as follows: due to the absence of quantum coherence, $\bar\rho^\alpha_{\bar t^-_{\rm m}}$ is diagonalized in the energy representation, i.e., $\bar\rho^\alpha_{\bar t^-_{\rm m}}=\sum_{n^*} \bar p^\alpha_{n^*}(\bar t_{\rm m})\Theta|n\rangle\langle n|\Theta^\dag$. Recalling that a general classical measurement operator takes the form $M_\alpha=\sum_n \sqrt{p_{\alpha|n}}|n\rangle\langle n|$, so that $\tilde M_\alpha=\sum_n \sqrt{p_{\alpha|n}}\Theta|n\rangle\langle n|\Theta^\dag$ and
\begin{equation}
\eta_{\rm QJT}=\sum_{\alpha,n} p_{\alpha|n}p^\alpha_{n^*}(\bar t_{\rm m})=\sum_\alpha \tilde p_{\alpha^*|\alpha}=\eta_{\rm C},
\end{equation}
where $p_{\alpha^*|\alpha}\equiv\sum_n p_{\alpha^*|n}p^\alpha_n(\bar t_{\rm m})$ and the symmetry $p_{\alpha^*|n^*}=p_{\alpha|n}$ has been assumed. Also, the system is assumed to be time-reversal invariant so that $\sum_n=\sum_{n^*}$, but it may have the Kramers degeneracy.

\subsection{Derivation of the relevant information gain (\ref{IG})}
\label{DIG}
To derive the second generalized Jarzynski equality, we again make use of Eq.~(\ref{DBTL}). Based on the definition $I_{\rm QJT}=\ln\|\tilde M_\alpha|\bar\psi_{\bar t^-_{\rm m}}\rangle\|^2-\ln p_\alpha$, we have
\begin{widetext}
\begin{equation}
\langle e^{-\beta(W-\Delta F)-I_{\rm QJT}}\rangle=\sum_\alpha\int D[\psi_t]\mathcal{P}[\psi_t,\alpha]e^{-\beta(W[\psi_t,\alpha]-\Delta F_\alpha)-I_{\rm QJT}[\psi_t,\alpha]}
=\sum_\alpha p_\alpha\int D[\bar \psi_t]\frac{\mathcal{\bar P}[\bar\psi_t,\alpha]}{\|\tilde M_\alpha|\bar\psi_{\bar t^-_{\rm m}}\rangle\|^2}.
\end{equation}
\end{widetext}
Each path integral $\int D[\bar \psi_t]\frac{\mathcal{\bar P}[\bar\psi_t,\alpha]}{\|\tilde M_\alpha|\bar\psi_{\bar t^-_{\rm m}}\rangle\|^2}$ on the last part of the above equation turns out to be unity (we set $\bar t_{\bar K_{\rm m}+1}\equiv \bar t_{\rm m}$ here for convenience):
\begin{widetext}
\begin{equation}
\begin{split}
&\sum_b\sum^\infty_{\bar K_{\rm m}=0}\sum_{\{\bar j_k:1\le k\le\bar K_{\rm m}\}}\prod^{\bar K_{\rm m}}_{k=1}\int^{\bar t_{k+1}}_0 d\bar t_k\; p^{\rm eq}_b(\bar\lambda^\alpha_0)\frac{\|\tilde M_\alpha \bar U^\alpha_{\mathrm{eff}}(\bar t_{\rm m},\bar t_{\bar K_{\rm m}}) \left[\prod^{\bar K_{\rm m}}_{k=1}\bar L_{\bar j_k}(\bar\lambda^\alpha_{\bar t_k})\bar U^\alpha_{\mathrm{eff}}(\bar t_k,\bar t_{k-1})\right] \Theta|b^{\bar\lambda^\alpha_0}\rangle\|^2}{\|\tilde M_\alpha|\bar\psi_{\bar t^-_{\rm m}}\rangle\|^2}\\
&=
\sum_b\sum^\infty_{\bar K_{\rm m}=0}\sum_{\{\bar j_k:1\le k\le\bar K_{\rm m}\}}\prod^{\bar K_{\rm m}}_{k=1}\int^{\bar t_{k+1}}_0 d\bar t_k\; p^{\rm eq}_b(\bar\lambda^\alpha_0)\|\bar U^\alpha_{\mathrm{eff}}(\bar t_{\rm m},\bar t_{\bar K_{\rm m}}) \left[\prod^{\bar K_{\rm m}}_{k=1}\bar L_{\bar j_k}(\bar\lambda^\alpha_{\bar t_k})\bar U^\alpha_{\mathrm{eff}}(\bar t_k,\bar t_{k-1})\right] \Theta|b^{\bar\lambda^\alpha_0}\rangle\|^2\\
&=\mathrm{Tr}[\mathcal{T}_+ e^{\int^{\bar t_{\rm m}}_0 dt \mathcal{\bar L}^\alpha_t}\bar\rho^{\rm eq}(\lambda^\alpha_\tau)]=1.
\end{split}
\end{equation}
\end{widetext}
Hence, we obtain 
\begin{equation}
\langle e^{-\beta(W-\Delta F)-I_{\rm QJT}}\rangle=\sum_\alpha p_\alpha=1.
\label{secondGJE}
\end{equation}
Here the explicit expression of $|\bar\psi_{\bar t^-_{\rm m}}\rangle$ 
\begin{widetext}
\begin{equation}
|\bar\psi_{\bar t^-_{\rm m}}\rangle=\frac{\bar U^\alpha_{\mathrm{eff}}(\bar t_{\rm m},\bar t_{\bar K_{\rm m}}) \left[\prod^{\bar K_{\rm m}}_{k=1}\bar L_{\bar j_k}(\bar\lambda^\alpha_{\bar t_k})\bar U^\alpha_{\mathrm{eff}}(\bar t_k,\bar t_{k-1})\right] \Theta|b^{\bar\lambda^\alpha_0}\rangle}{\|\bar U^\alpha_{\mathrm{eff}}(\bar t_{\rm m},\bar t_{\bar K_{\rm m}}) \left[\prod^{\bar K_{\rm m}}_{k=1}\bar L_{\bar j_k}(\bar\lambda^\alpha_{\bar t_k})\bar U^\alpha_{\mathrm{eff}}(\bar t_k,\bar t_{k-1})\right] \Theta|b^{\bar\lambda^\alpha_0}\rangle\|},
\end{equation}
\end{widetext}
has been used. Accordingly, the ket can be expressed in terms of the measurement outcomes in the forward QJT
\begin{widetext}
\begin{equation}
\begin{split}
\langle\bar\psi_{\bar t^-_{\rm m}}|=\frac{\langle b^{\bar\lambda^\alpha_0}|\Theta^\dag\left[\prod^1_{k=\bar K_{\rm m}}\bar U^{\alpha\dag}_{\mathrm{eff}}(\bar t_k,\bar t_{k-1})\bar L^\dag_{\bar j_k}(\bar\lambda^\alpha_{\bar t_k})\right]\bar U^{\alpha\dag}_{\mathrm{eff}}(\bar t_{\rm m},\bar t_{\bar K_{\rm m}})}{\|\langle b^{\bar\lambda^\alpha_0}|\Theta^\dag\left[\prod^1_{k=\bar K_{\rm m}}\bar U^{\alpha\dag}_{\mathrm{eff}}(\bar t_k,\bar t_{k-1})\bar L^\dag_{\bar j_k}(\bar\lambda^\alpha_{\bar t_k})\right]\bar U^{\alpha\dag}_{\mathrm{eff}}(\bar t_{\rm m},\bar t_{\bar K_{\rm m}})\|}\\
=\frac{\langle b^{\lambda^\alpha_\tau}|\left[\prod^K_{k=K_{\rm m}+1} U^\alpha_{\mathrm{eff}}(t_{k+1},t_k)L_{j_k}(\lambda^\alpha_{t_k})\right] U^{\alpha}_{\mathrm{eff}}(t_{K_{\rm m}+1},t_{\rm m})\Theta^\dag}{\|\langle b^{\lambda^\alpha_\tau}|\left[\prod^K_{k=K_{\rm m}+1} U^\alpha_{\mathrm{eff}}(t_{k+1},t_k)L_{j_k}(\lambda^\alpha_{t_k})\right] U^{\alpha}_{\mathrm{eff}}(t_{K_{\rm m}+1},t_{\rm m})\|},
\end{split}
\end{equation}
\end{widetext}
where $L^\dag_{\bar j_k}(\bar\lambda_{\bar t_k})\propto L_{j_{K+1-k}}(\lambda_{t_{K+1-k}})$ has been used. One can see that generally all the measurement outcomes after $t_{\rm m}$ should be used to determine $|\bar\psi_{\bar t^-_{\rm m}}\rangle$, which usually differs from $\Theta|\psi_{t^+_{\rm m}}\rangle$. 
One can also see that the validity of the second generalized Jarzynski equality only requires $\sum_\alpha p_\alpha=1$, so they are not necessarily the real probabilities of the measurement outcomes. However, to minimize the averaged value $\langle I_{\rm QJT}\rangle$, which gives an upper bound of $-\beta\langle W_{\rm diss}\rangle$, the real measurement outcome probabilities are the optimal choice. Another advantage is that, $\langle I_{\rm QJT}\rangle$ has a Holevo bound-like expression under such a choice.

To measure $I_{\rm QJT}$, we have to measure both $\|\tilde M_\alpha|\bar\psi_{\bar t^-_{\rm m}}\rangle\|^2$ and $p_\alpha$. The latter is straightforward since we have only to count the number of all the possible measurement outcomes, and then perform the statistical estimation after many repeats of the feedback control experiment. On the other hand, measuring $\|M_\alpha|\bar\psi_{\bar t^-_{\rm m}}\rangle\|^2$ is, though in principle feasible, much more involved: for given $\alpha$, we should prepare a sufficiently large amount of realizations of the time-reversed processes to observe, under a certain coarse graining of time, all the possible outcomes $dN^j_t$ from monitoring the heat bath. Conditioned on each sequence of outcomes, we perform the measurement $\mathrm{M}_{\mathrm{B}_\alpha}$ to statistically determine the conditional probability $\|\tilde M_\alpha|\bar\psi_{\bar t^-_{\rm m}}\rangle\|^2$, which again requires many repetitions. Fortunately, if there are only state transition QJs, we can simplify the above process into the following procedure: for given $\alpha$, we start from the $b$-th instantaneous energy eigenstate of $\bar H(\bar\lambda_t)$ at different times $t>t_{\rm m}$ and apply the time-reversed driving protocol $\bar\lambda^\alpha_t$ and $\bar h^\alpha_t$. We then perform the measurement $\mathrm{M}_{\mathrm{B}_\alpha}$ to estimate the conditional probability $\tilde p_{\alpha|b,t}$ of that outcome $\alpha$ being observed for those QJTs with no QJ after $t-t_{\rm m}$. The probability $\tilde p_{\alpha|b,t}$ has already covered all the possible $\|\tilde M_\alpha|\bar\psi_{\bar t^-_{\rm m}}\rangle\|^2$. This fact may be accounted for by the completely destructive nature of a state-transition QJ (or a PM performed at the final stage) that makes all measurement outcomes after $t_{\rm m}$ irrelevant to estimate the quantum state at $t^+_{\rm{m}}$, and this fact has been used in our numerical calculations. One can also see that the knowledge of the microscopic details about the system and the measurement is not needed in a real experiment - we only have to deal with the classical outcomes.

The consistency between $I_{\rm QJT}$ and the classical mutual information $I_{\rm C}$ at the trajectory level is transparent: in the classical limit, we have $|\bar\psi_{\bar t^-_{\rm m}}\rangle=\Theta|\psi_{t_{\rm m}}\rangle$ with $|\psi_{t_{\rm m}}\rangle$ being a certain eigenstate $|n_t\rangle$. Recalling the general classical form of $M_\alpha$, we have 
\begin{equation}
\begin{split}
I_{\rm QJT}[\psi_t,\alpha]&=\ln\|\Theta M_\alpha|n_t\rangle\|^2-\ln p_\alpha\\
&=\ln p_{\alpha|n_t}-\ln p_\alpha=I_{\rm C}[n_t,\alpha].
\end{split}
\end{equation}

\subsection{Derivation of Eq.~(\ref{AIG}) and the properties of relevant information}
\label{DAIG}
By definition, the average value of $I_{\rm QJT}$ should be
\begin{widetext}
\begin{equation}
\begin{split}
&\langle I^{\rm QJT}\rangle=\sum_\alpha\int D[\psi_t]\mathcal{P}[\psi_t,\alpha] I^{\rm QJT}[\psi_t,\alpha]\\
&=\sum_{\alpha,b}\sum^\infty_{K=K_{\rm m}}\sum_{\{j_k:K_{\rm m}< k\le K\}}\prod^K_{k=K_{\rm m}+1}\int^{t_{k+1}}_{t_{\rm m}} dt_k\|\langle b^{\lambda^\alpha_\tau}|\left[\prod^K_{k=K_{\rm m}+1} U^\alpha_{\mathrm{eff}}(t_{k+1},t_k)L_{j_k}(\lambda^\alpha_{t_k})\right] U^{\alpha}_{\mathrm{eff}}(t_{K_{\rm m}+1},t_{\rm m})M_\alpha\sqrt{\rho_{t^-_{\rm m}}}\|^2\\
&\times\ln \frac{\|\langle b^{\lambda^\alpha_\tau}|\left[\prod^K_{k=K_{\rm m}+1} U^\alpha_{\mathrm{eff}}(t_{k+1},t_k)L_{j_k}(\lambda^\alpha_{t_k})\right] U^{\alpha}_{\mathrm{eff}}(t_{K_{\rm m}+1},t_{\rm m})M_\alpha\|^2}{p_\alpha\|\langle b^{\lambda^\alpha_\tau}|\left[\prod^K_{k=K_{\rm m}+1} U^\alpha_{\mathrm{eff}}(t_{k+1},t_k)L_{j_k}(\lambda^\alpha_{t_k})\right] U^{\alpha}_{\mathrm{eff}}(t_{K_{\rm m}+1},t_{\rm m})\|^2}.
\end{split}
\label{AIQJT}
\end{equation}
\end{widetext}
Based on the definition $\mathcal{I}_{\rm C}(\rho:\mathrm{M_X})\equiv H(p^{\mathrm{M_X}}_\rho||p^{\mathrm{M_X}}_{\rho_{\rm u}})$, we can write down
\begin{widetext}
\begin{equation}
\begin{split}
&\mathcal{I}_{\rm C}(\rho^\alpha_{t^+_{\rm m}}:\Pi^{\lambda^\alpha_\tau}\mathrm{M}_{J_{t_{\rm m}<t<\tau|\alpha}})=\\
&\sum_b\sum^\infty_{K=K_{\rm m}}\sum_{\{j_k:K_{\rm m}< k\le K\}}\prod^K_{k=K_{\rm m}+1}\int^{t_{k+1}}_{t_{\rm m}} dt_k\|\langle b^{\lambda^\alpha_\tau}|\left[\prod^K_{k=K_{\rm m}+1} U^\alpha_{\mathrm{eff}}(t_{k+1},t_k)L_{j_k}(\lambda^\alpha_{t_k})\right] U^{\alpha}_{\mathrm{eff}}(t_{K_{\rm m}+1},t_{\rm m})\sqrt{\rho^\alpha_{t^+_{\rm m}}}\|^2\\
&\times\ln \frac{\|\langle b^{\lambda^\alpha_\tau}|\left[\prod^K_{k=K_{\rm m}+1} U^\alpha_{\mathrm{eff}}(t_{k+1},t_k)L_{j_k}(\lambda^\alpha_{t_k})\right] U^{\alpha}_{\mathrm{eff}}(t_{K_{\rm m}+1},t_{\rm m})\sqrt{\rho^\alpha_{t^+_{\rm m}}}\|^2}{\|\langle b^{\lambda^\alpha_\tau}|\left[\prod^K_{k=K_{\rm m}+1} U^\alpha_{\mathrm{eff}}(t_{k+1},t_k)L_{j_k}(\lambda^\alpha_{t_k})\right] U^{\alpha}_{\mathrm{eff}}(t_{K_{\rm m}+1},t_{\rm m})\sqrt{\rho_{\rm u}}\|^2},\\
&\mathcal{I}_{\rm C}(\rho_{t^-_{\rm m}}:\Pi^{\lambda^{\rm A}_\tau}\mathrm{M}_{J_{t_{\rm m}<t<\tau|\rm A}}\mathrm{M_A})=\\
&\sum_{\alpha,b}\sum^\infty_{K=K_{\rm m}}\sum_{\{j_k:K_{\rm m}< k\le K\}}\prod^K_{k=K_{\rm m}+1}\int^{t_{k+1}}_{t_{\rm m}} dt_k\|\langle b^{\lambda^\alpha_\tau}|\left[\prod^K_{k=K_{\rm m}+1} U^\alpha_{\mathrm{eff}}(t_{k+1},t_k)L_{j_k}(\lambda^\alpha_{t_k})\right] U^{\alpha}_{\mathrm{eff}}(t_{K_{\rm m}+1},t_{\rm m})M_\alpha\sqrt{\rho_{t^-_{\rm m}}}\|^2\\
&\times\ln \frac{\|\langle b^{\lambda^\alpha_\tau}|\left[\prod^K_{k=K_{\rm m}+1} U^\alpha_{\mathrm{eff}}(t_{k+1},t_k)L_{j_k}(\lambda^\alpha_{t_k})\right] U^{\alpha}_{\mathrm{eff}}(t_{K_{\rm m}+1},t_{\rm m})M_\alpha\sqrt{\rho_{t^-_{\rm m}}}\|^2}{\|\langle b^{\lambda^\alpha_\tau}|\left[\prod^K_{k=K_{\rm m}+1} U^\alpha_{\mathrm{eff}}(t_{k+1},t_k)L_{j_k}(\lambda^\alpha_{t_k})\right] U^{\alpha}_{\mathrm{eff}}(t_{K_{\rm m}+1},t_{\rm m})M_\alpha\sqrt{\rho_{\rm u}}\|^2}.
\end{split}
\label{IC}
\end{equation}
\end{widetext}
Combining Eq.~(\ref{IC}) with Eq.~(\ref{AIQJT}), using $\rho^\alpha_{t^+_{\rm m}}\equiv M_\alpha\rho_{t^-_{\rm m}}M^\dag_\alpha/p_\alpha$, we finally obtain
\begin{equation}
\begin{split}
\langle I^{\rm QJT}\rangle&=\sum_\alpha p_\alpha\mathcal{I}_{\rm C}(\rho^\alpha_{t^+_{\rm m}}:\Pi^{\lambda^\alpha_\tau}\mathrm{M}_{J_{t_{\rm m}<t<\tau|\alpha}})\\
&-\mathcal{I}_{\rm C}(\rho_{t^-_{\rm m}}:\Pi^{\lambda^{\rm A}_\tau}\mathrm{M}_{J_{t_{\rm m}<t<\tau|\rm A}}\mathrm{M_A}).
\end{split}
\end{equation}

The fact that $\mathcal{I}_{\rm C}$ is always bounded by $\mathcal{I}_{\rm Q}$ can be understood intuitively: $\mathcal{I}_{\rm Q}(\rho)$ is the intrinsic information that the quantum state $\rho$ carries, while $\mathcal{I}_{\rm C}(\rho:\rm{M_X})$ is the available information content extracted from the classical outcomes by a measurement $\rm{M_X}$ performed on $\rho$. This result can be obtained from the following relation \cite{Barchielli2005}:
\begin{equation}
\begin{split}
&S(\rho||\sigma)\ge S\left(\bigoplus_x M_x\rho M^\dag_x||\bigoplus_x M_x\sigma M^\dag_x\right)\\
&=H(p^{\rm M_X}_\rho||p^{\rm M_X}_\sigma)+\sum_x p^x_\rho S(\rho_x||\sigma_x)\ge H(p^{\rm M_X}_\rho||p^{\rm M_X}_\sigma),
\end{split}
\end{equation}
where $\rho_x\equiv M_x\rho M^\dag_x/p^x_\rho$ and $p^x_\rho=\mathrm{Tr}[M_x\rho M^\dag_x]$ for $x\in\rm{X}$. Another good property of $\mathcal{I}_{\rm C}$ is that it increases monotonically when performing subsequent measurements, namely $\mathcal{I}_{\rm C}(\rho:\rm{M_YM_X})\ge\mathcal{I}_{\rm C}(\rho:\rm{M_X})$. This is a result of the chain rule of the classical relative entropy \cite{Cover2006}:
\begin{equation}
\begin{split}
&H(p^{\rm M_YM_X}_\rho||p^{\rm M_YM_X}_\sigma)=H(p^{\rm M_X}_\rho||p^{\rm M_X}_\sigma)\\
&+\sum_x p^x_\rho H(p^{\rm M_Y}_{\rho_x}||p^{\rm M_Y}_{\sigma_x})\ge H(p^{\rm M_X}_\rho||p^{\rm M_X}_\sigma).
\end{split}
\end{equation}
From the above equation we can also find that $\mathcal{I}_{\rm C}(\rho:\rm{M_YM_X})=\mathcal{I}_{\rm C}(\rho:\rm{M_X})$ once $\rm M_Y$ is a projective measurement, no matter how complex $\rm M_X$ is (e.g., a combination of $\mathrm{M}_{\mathrm{X}_k}$).

\subsection{Other fluctuation theorems}
\label{OFT}
In a real quantum feedback control experiment, we only perform the initial and the final PMs to determine the energy change, a general measurement $\mathrm{M}_{\mathrm{A}}$ for feedback, and the continuous monitoring of the heat bath to determine the heat along a single trajectory. Our results in the main text are fully compatible with such an experiment, and the correction term $I[\psi_t,\alpha]$ is in principle measurable. However, if we only concern the ensemble average value, we can insert arbitrary numbers of nondemolition PMs \footnote{To perform a nondemolition PM on $\rho$, we explicitly mean that the basis of the PM is exactly the eigenbasis of $\rho$. Therefore, such ``nondemolition" PM actually disturbes the system at the $\mathcal{E}_P$-ensemble level \cite{Breuer2002} and thus affects the work or heat distributions, despite it has no backaction at the $\mathcal{E}_\rho$-ensemble level and thus preserves $\langle W\rangle$ or $\langle Q\rangle$.} at arbitrary time points while keeping $\langle W\rangle$ ($\langle Q\rangle$ or $\langle\Delta s\rangle$) unchanged, since a nondemolition PM preserves the density operator and costs no work (but does affect the work and heat fluctuations). Such a technique was used in Ref.~\cite{Esposito2006} to construct a classical trajectory (\ref{CT})-like quantum trajectory (continuously perform nondemolition PMs on the system), though the experimental realization is difficult. Particularly, if we insert two nondemolition PMs right before and after $\mathrm{M}_{\mathrm{A}}$, we can still construct the same second-type generalized Jarzynski equality (\ref{secondGJE}) in form by redefining the correction term $I_{\rm QJT}$ as \cite{Funo2013}
\begin{equation}
I_{\mathrm{ba}}=\ln p_{l|\alpha}-\ln p_k,
\end{equation}
where $\rho_{t^-_\mathrm{m}}=\sum_k p_k|k\rangle\langle k|$ and $\rho^\alpha_{t^+_\mathrm{m}}=\sum_l p_{l|\alpha}|l^\alpha\rangle\langle l^\alpha|$. After taking the ensemble average, we obtain the QC-mutual information \cite{Sagawa2008}:
\begin{equation}
\langle I_{\mathrm{ba}}\rangle=\sum_\alpha p_\alpha\mathcal{I}_{\mathrm{Q}}(\rho^\alpha_{t^+_{\mathrm{m}}})-\mathcal{I}_{\mathrm{Q}}(\rho_{t^-_{\mathrm{m}}})=I_{\mathrm{QC}}.
\end{equation}
Although the feedback control process compatible with $I_{\rm ba}$ is somehow artificial, 
$I_{\mathrm{QC}}$ indeed gives an upper bound for $-\beta\langle W_{\rm diss}\rangle$ in real experiments (without the two nondemolition PMs) at the ensemble level.

In fact, we have two other second-type generalized Jarzynski equalities, which respectively correspond to the feedback control processes with \emph{only one} nondemolition PM just before \emph{or} after $\mathrm{M}_{\mathrm{A}}$. Concretely, for the case of a PM immediately before $\rm M_A$, we define $I_{\mathrm{b}}$ as
\begin{equation}
I_{\mathrm{b}}=\ln\langle\bar\psi_{\bar t^-_m}|\Theta\rho^\alpha_{t^+_m}\Theta^\dagger|\bar\psi_{\bar t^-_m}\rangle-\ln p_k,
\end{equation}
for which the ensemble average is
\begin{equation}
\langle I_{\mathrm{b}}\rangle=\sum_\alpha p_\alpha\mathcal{I}_{\mathrm{C}}(\rho^\alpha_{t^+_{\mathrm{m}}}:\Pi_{\tau|\alpha}\mathrm{M}_{J_{t_m<t<\tau}|\alpha})-\mathcal{I}_{\mathrm{Q}}(\rho_{t^-_\mathrm{m}}).
\end{equation}
Recalling that $\mathcal{I}_{\rm C}$ is always bounded by $\mathcal{I}_{\rm Q}$, this bound is always tighter than \emph{both} $\langle I_{\rm{QJT}}\rangle$ \emph{and} $I_{\mathrm{QC}}$. If we want to saturate $\langle I_{\mathrm{b}}\rangle$ to $I_{\mathrm{QC}}$, the only chance for it seems to first quench the Hamiltonian to commute with $\rho_{t^+_{\mathrm{m}}}$ followed by a quasistatic process, as proposed in Ref.~\cite{Jacobs2009}. For the case of a PM just after $\rm M_A$, we define $I_{\mathrm{a}}$ as
\begin{equation}
I_{\mathrm{a}}=\ln \|M^\dagger_\alpha |l_\alpha\rangle\|^2-\ln p_\alpha,
\end{equation}
of which the ensemble average is
\begin{equation}
\langle I_{\mathrm{a}}\rangle=\sum_\alpha p_\alpha\mathcal{I}_{\mathrm{Q}}(\rho^\alpha_{t^+_{\mathrm{m}}})-\mathcal{I}_{\mathrm{C}}(\rho_{t^-_{\mathrm{m}}}:\Pi^{\rm nd}_{t^+_{\rm m}|\mathrm{A}} \mathrm{M}_{\mathrm{A}}),
\end{equation}
where $\Pi^{\rm nd}_{t^+_{\rm m}|\alpha}\equiv\{|l_\alpha\rangle\langle l_\alpha|:l=1,2,...,d\}$ is the nondemolition PM with respect to $\rho^\alpha_{t^+_{\rm m}}$. This bound is the loosest compared with the other three bounds.

\section{Details of the example}
\label{Example}
\subsection{Equation of motion (\ref{TLSEOM})}
\label{EOME}
The equation of motion used in the ``example" part, with the external driving turned off, can be obtained from the following standard total Hamiltonian \cite{Breuer2002,Wiseman2010}:
\begin{equation}
\begin{split}
H_{\mathrm{tot}}(t)&=\frac{1}{2}\hbar\omega_t\sigma_z\otimes I_{\mathrm{B}}+I_{\mathrm{S}}\otimes \sum_{\boldsymbol{k}}\hbar\omega_{\boldsymbol{k}}b^\dag_{\boldsymbol{k}}b_{\boldsymbol{k}} \\
&+g\sigma_x\otimes \sum_{\boldsymbol{k}}(c_{\boldsymbol{k}}b^\dag_{\boldsymbol{k}}+c^*_{\boldsymbol{k}}b_{\boldsymbol{k}}),
\end{split}
\end{equation}
where a two-level system is coupled to a noninteracting many-boson heat bath. The first condition in Eq.~(\ref{judgement}) holds true due to $q(\omega)\equiv0$ (here $H(\omega)=\hbar\omega\sigma_z/2$). When the heat bath is at equilibrium, the correlation function can be obtained as
\begin{equation}
\mathcal{B}(t)=\sum_{\boldsymbol{k}} |c_{\boldsymbol{k}}|^2[\langle n_{\boldsymbol{k}}\rangle e^{i\omega_{\boldsymbol{k}}t}+(\langle n_{\boldsymbol{k}}\rangle+1)e^{-i\omega_{\boldsymbol{k}}t}],
\end{equation}
where $\langle n_{\boldsymbol{k}}\rangle=(e^{\beta\hbar\omega_{\boldsymbol{k}}}-1)^{-1}$ and the Fourier transform of $\mathcal{B}(t)$, denoted by $\Gamma(\omega)=\int^{+\infty}_{-\infty}dt e^{i\omega t}\mathcal{B}(t)$, reads 
\begin{equation}
\Gamma(\omega)=\frac{1}{1-e^{-\beta\hbar\omega}}[J(\omega)-J(-\omega)],
\end{equation}
where $J(\omega)\equiv2\pi\sum_{\boldsymbol{k}}|c_{\boldsymbol{k}}|^2\delta(\omega-\omega_{\boldsymbol{k}})$ is the spectral function. After assuming an Ohmic spectral $J(\omega)=\kappa_0\omega\theta(\omega)$ ($\theta(\omega)$: Heaviside step function), we find that the only two nonvanishing jump operators are $L_{\pm}(\omega_t)=\sqrt{\gamma_\pm(\omega_t)}\sigma_\pm$, $\sigma_\pm=(\sigma_x\pm i\sigma_y)/2$, where the transition rates read 
\begin{equation}
\gamma_{\pm}(\omega)=\frac{g^2}{\hbar^2}\Gamma(\mp\omega)=\frac{1}{2}\kappa\omega\left(\coth\frac{\beta\hbar\omega}{2}\mp1\right),
\end{equation}
with $\kappa\equiv\kappa_0\frac{g^2}{\hbar^2}$. The memory time $\tau_{\rm B}$ is of the order of $\beta\hbar$ \cite{Lidar2012}, so the second condition in Eq.~(\ref{judgement}) becomes $\beta g\ll1$ and $\beta g^2\ll \min_{0\le t\le\tau}\omega_t$, which is well satisfied for the parameters we use ($\beta=5,g=0.001$ and $\omega_0=0.3$). If we neglect the Lamb shift, we will obtain the following adiabatic Lindblad equation
\begin{equation}
\dot\rho_t=-\frac{i}{2}[\omega_t\sigma_z,\rho_t]+\sum_{j=\pm}\gamma_j(\omega_t)\mathcal{D}[\sigma_j]\rho_t,
\end{equation}
which gives Eq.~(\ref{TLSEOM}) in the main text by further adding the perturbative driving term $\epsilon\sigma_x\cos\omega_{\rm d}t$.

\subsection{Numerical calculations}
\label{NC}
We apply the standard stochastic wave function approach \cite{Dalibard1992}, as was used in Ref.~\cite{Pekola2013}. We analyze a total of $10^6$ individual QJTs. For the feedback control process in a single realization, we first  generate a random number $X$ which distributes uniformly over $[0,1]$ ($X\sim U[0,1]$) for initialization. If $X<p_e(\omega_0)=e^{-\frac{\beta\hbar\omega_0}{2}}/(2\cosh\frac{\beta\hbar\omega_0}{2})$, we initialize the system as $|\psi_0\rangle=|e\rangle$, record the initial energy $E_i=\hbar\omega_0/2$ and set the coherent driving strength $\epsilon=0.008$. Otherwise, we have $|\psi_0\rangle=|g\rangle$, $E_i=-\hbar\omega_0/2$ and $\epsilon=0.002$ ($\sigma_z|e\rangle=|e\rangle$ and $\sigma_z|g\rangle=-|g\rangle$). For the corresponding ordinary process (without feedback control), we always set $\epsilon=0.008p_e(\omega_0)+0.002p_g(\omega_0)=0.0031$ whatever the initial state is. The accumulated heat $Q$ is initialized to $0$.

We discretize the time interval $[0,\tau=2000]$ into 20000 identical parts \footnote{The results are almost unchanged even if we double the total step numbers. Similar observations have been highlighted in Ref.~\cite{Pekola2013}.}, each with length $\Delta t=0.1$. For each time step, we use $e^{-\frac{i}{\hbar}H_{\rm eff}(t+\Delta t/2)\Delta t}$ to approximate the effective time-evolution operator $U_{\rm eff}(t+\Delta t,t)$, where
\begin{equation}
\begin{split}
H_{\rm eff}(t)&=\frac{\hbar}{2}(\omega_t\sigma_z+\epsilon\sigma_x\cos\omega_{\rm d}t)\\
&-\frac{i\hbar}{4}\kappa\omega_t\left(\sigma_z+\coth\frac{\beta\hbar\omega_t}{2}\right).
\end{split}
\label{Heff}
\end{equation}
Suppose that the system state is $|\psi_t\rangle=c_e(t)|e\rangle+c_g(t)|g\rangle$ at time $t$. To determine the state at $t+\Delta t$, we should first calculate
\begin{equation}
\Delta p=1-\|e^{-\frac{i}{\hbar}H_{\rm eff}(t+\Delta t/2)\Delta t}|\psi_t\rangle\|^2,
\end{equation} 
which is the probability that a QJ occurs. To make the event probabilistic, we generate a random number $Y_t\sim U[0,1]$. If $Y_t>\Delta p$, the time evolution is determined by
\begin{equation}
|\psi_{t+\Delta t}\rangle=\frac{e^{-\frac{i}{\hbar}H_{\rm eff}(t)\Delta t}|\psi_t\rangle}{\sqrt{1-\Delta p}}.
\end{equation}
Otherwise, one of the two possible QJs occurs. The ratio of the probabilities between a de-excitation QJ and an excitation QJ is $|c_e(t)|^2\gamma_-(\omega_t)/|c_g(t)|^2\gamma_+(\omega_t)$. Therefore, we independently generate another random variable $Z_t\sim U[0,1]$. If $Z_t<\frac{|c_e(t)|^2\gamma_-(\omega_t)}{|c_e(t)|^2\gamma_-(\omega_t)+|c_g(t)|^2\gamma_+(\omega_t)}$, a de-excitation QJ occurs, so that
\begin{equation}
|\psi_{t+\Delta t}\rangle=|g\rangle,
\end{equation}
and the accumulated heat increases by $dQ=\hbar\omega_t$. Otherwise, an excitation QJ occurs, so that 
\begin{equation}
|\psi_{t+\Delta t}\rangle=|e\rangle,
\end{equation}
and $dQ=-\hbar\omega_t$. Finally, we projectively measure $|\psi_{\tau^-}\rangle$ under the basis $\{|e\rangle,|g\rangle\}$, with probability $|\langle e|\psi_{\tau^-}\rangle|^2$ ($|\langle g|\psi_{\tau^-}\rangle|^2$) to observe outcome $e$ ($g$). To this end, we generate a random variable $X'\sim U[0,1]$. If $X'<|\langle e|\psi_{\tau^-}\rangle|^2$, we record the final energy $E_f=\hbar\omega_\tau/2$. Otherwise, we have $E_f=-\hbar\omega_\tau/2$. Now the work during this single run can be evaluated as
\begin{equation}
W=E_f-E_i+Q.
\end{equation}

Now $\eta_{\rm QJT}$ is obtained by numerically solving the time-reversed LME in the $\sigma_z$ representation
\begin{widetext}
\begin{displaymath}
\frac{d}{dt}\left[\begin{array}{c}\rho_{ee}(t)\\ \rho_{gg}(t)\\ \rho_{eg}(t)\\ \rho_{ge}(t)\end{array}
\right] =
\left[ \begin{array}{cccc}
-\gamma_-(\bar\omega_t) & \gamma_+(\bar\omega_t) & \frac{i}{2}\epsilon\cos\omega_{\rm d}t & -\frac{i}{2}\epsilon\cos\omega_{\rm d}t \\
\gamma_-(\bar\omega_t) & - \gamma_+(\bar\omega_t) & -\frac{i}{2}\epsilon\cos\omega_{\rm d}t & \frac{i}{2}\epsilon\cos\omega_{\rm d}t  \\
\frac{i}{2}\epsilon\cos\omega_{\rm d}t  & -\frac{i}{2}\epsilon\cos\omega_{\rm d}t & -\frac{\gamma_+(\bar\omega_t)+\gamma_-(\bar\omega_t)}{2}-i\bar\omega_t & 0 \\
-\frac{i}{2}\epsilon\cos\omega_{\rm d}t  & \frac{i}{2}\epsilon\cos\omega_{\rm d}t & 0 & -\frac{\gamma_+(\bar\omega_t)+\gamma_-(\bar\omega_t)}{2}+i\bar\omega_t  \\
\end{array} \right]
\left[\begin{array}{c}\rho_{ee}(t)\\ \rho_{gg}(t)\\ \rho_{eg}(t)\\ \rho_{ge}(t)\end{array}\right],
\end{displaymath}
\end{widetext}
where $\bar\omega_t=\omega_{\tau-t}=\omega_\tau-\Delta\omega t/\tau$, $\omega_\tau=\omega_0+\Delta\omega$ and the symmetry $h_{\tau-t}=h_t$ has been used. The initial condition is the equilibrium state at $\omega=\omega_\tau$, i.e., $\rho_{ee}(0)=e^{-\frac{\beta\hbar\omega_\tau}{2}}/(2\cosh\frac{\beta\hbar\omega_\tau}{2})=1-\rho_{gg}(0)$ and $\rho_{eg}(0)=\rho_{ge}(0)=0$. For $\alpha=g$ ($\alpha=e$), we solve the above equation with $\epsilon=0.002$ ($\epsilon=0.008$) to obtain the final density matrix, so that $\tilde p_e=\rho_{ee}(\tau)$ ($\tilde p_g=\rho_{gg}(\tau)$). Finally, we obtain $\eta_{\rm QJT}=\tilde p_e+\tilde p_g$.

As mentioned in Appendix \ref{DIG}, the two ingredients to determine $I_{\rm QJT}$ are $p_\alpha$ and $\|\tilde M_\alpha|\bar\psi_{\bar t^-_{\rm m}}\rangle\|^2$. Here $p_\alpha$ simply equals the initial canonical distribution of state $\alpha$ ($e$ or $g$). While it is generally difficult to calculate $\|\tilde M_\alpha|\bar\psi_{\bar t^-_{\rm m}}\rangle\|^2$, without dephasing QJs, it can be obtained from (i) the information (channel index and time) of the first QJ after $t_{\rm m}$, or (ii) the final PM outcome if no QJ occurs during $[t_{\rm m},\tau]$. In particular, in our two-level model with $t_{\rm m}=0$, given the initial PM outcome $\alpha$ and the first QJ from the state $x$ (either $e$ or $g$) to the other state at $t_1=\tau-\bar t_K$, we have
\begin{equation}
\|\tilde M_\alpha|\bar\psi_{\bar t^-_{\rm m}}\rangle\|^2=\frac{|\langle \alpha|\bar U_{\rm eff}(\tau,\bar t_K)|x\rangle|^2}{\|\bar U_{\rm eff}(\tau,\bar t_K)|x\rangle\|^2},
\end{equation}
where $U_{\rm eff}(t_2,t_1)=\mathcal{T}_+ e^{-\frac{i}{\hbar}\int^{t_2}_{t_1} H_{\rm eff}(\tau-t)dt}$ with $H_{\rm eff}(t)$ given by Eq.~(\ref{Heff}). Finally, $\langle I_{\rm QJT}\rangle$ is obtained by taking the average over all these $I_{\rm QJT}$ data ($10^6$ in total).

\bibliography{GZP_references}

\end{document}